\documentclass[nonacm]{acmart} 

\usepackage{graphicx}%
\usepackage{multirow}%
\usepackage{amsmath,amsfonts}%
\usepackage{amsthm}%
\usepackage{mathrsfs}%
\usepackage[title]{appendix}%
\usepackage{xcolor}%
\usepackage{textcomp}%
\usepackage{manyfoot}%
\usepackage{booktabs}%
\usepackage{algorithm}%
\usepackage{algorithmicx}%
\usepackage{algpseudocode}%
\usepackage{listings}%
\usepackage{subfig}
\usepackage{hyperref}
\usepackage{color}
\usepackage[normalem]{ulem}
\usepackage{tabularray}
\usepackage{hyperref}
\usepackage{tablefootnote}
\usepackage{tabularx}
\usepackage{threeparttable}
\usepackage{anyfontsize}
\usepackage[export]{adjustbox}

\definecolor{gray}{HTML}{808080}
\definecolor{teal}{HTML}{21908C}
\definecolor{yellow}{HTML}{FDE725}
\definecolor{blue}{HTML}{1E2BC5}
\definecolor{purple}{HTML}{440154}
\definecolor{green}{HTML}{238E12}

\newtoggle{comments}
\toggletrue{comments}

\iftoggle{comments}{
  \newcommand{\notejl}[1]
  {{\color{teal}[{\bf Jinsook:} #1]}}
  \newcommand{\noterk}[1]
  {{\color{orange}[{\bf Rene:} #1]}}
    \newcommand{\noteeh}[1]
  {{\color{magenta}[{\bf Emma:} #1]}}
    \newcommand{\notejz}[1]
  {{\color{teal}[{\bf JZ:} #1]}}
    
    \newcommand{\notetj}[1]
  {{\color{blue}[{\bf Thorsten:} #1]}}
  \newcommand{\todo}[1]
  {{\color{red}[{\bf Todo:} #1]}} 
  \newcommand{\cut}[1]
  {{\color{red}\sout{#1}}}

}{
  \newcommand{\notejl}[1]{}
  \newcommand{\noterk}[1]{}
  \newcommand{\noteeh}[1]{}
        \newcommand{\notejz}[1]{}
        \newcommand{\noteng}[1]{}
        \newcommand{\notetj}[1]{}
  \newcommand{\todo}[1]{}
  \newcommand{\cut}[1]{}

}

\begin{document}

\title{Algorithms for College Admissions Decision Support: Impacts of Policy Change and Inherent Variability}

\author{Jinsook Lee}
\authornote{Both authors contributed equally to this research.}
\email{jl3369@cornell.edu}
\affiliation{%
  \institution{Cornell University}
  \city{Ithaca}
  \state{New York}
  \country{USA}}

\author{Emma Harvey}
\authornotemark[1]
\orcid{0000-0001-8453-4963}
\email{evh29@cornell.edu}
\affiliation{%
  \institution{Cornell University}
  \city{Ithaca}
  \state{New York}
  \country{USA}}

\author{Joyce Zhou}
\email{jz549@cornell.edu}
\affiliation{%
  \institution{Cornell University}
  \city{Ithaca}
  \state{New York}
  \country{USA}}

\author{Nikhil Garg}
\email{ngarg@cornell.edu}
\affiliation{%
  \institution{Cornell Tech}
  \city{New York}
  \state{New York}
  \country{USA}}

\author{Thorsten Joachims}
\email{tj@cs.cornell.edu}
\affiliation{%
  \institution{Cornell University}
  \city{Ithaca}
  \state{New York}
  \country{USA}}

\author{Rene F. Kizilcec}
\email{kizilcec@cornell.edu}
\affiliation{%
  \institution{Cornell University}
  \city{Ithaca}
  \state{New York}
  \country{USA}}

\renewcommand{\shortauthors}{Lee, Harvey, Zhou, Garg, Joachims, and Kizilcec}

\begin{abstract}
Each year, selective American colleges sort through tens of thousands of applications to identify a first-year class that displays both \textbf{academic merit} and \textbf{diversity}. In the 2023-2024 admissions cycle, these colleges faced unprecedented challenges to doing so. First, the \textbf{number of applications} has been steadily growing year-over-year. Second, \textbf{test-optional policies} that have remained in place since the COVID-19 pandemic limit access to key information that has historically been predictive of academic success. Most recently, longstanding debates over affirmative action culminated in the Supreme Court \textbf{banning race-conscious admissions}. Colleges have explored machine learning (ML) models to address the issues of scale and missing test scores, often via ranking algorithms intended to allow human reviewers to focus attention on `top' applicants. However, the Court's ruling will force changes to these models, which were previously able to consider race as a factor in ranking. There is currently a poor understanding of how these mandated changes will shape applicant ranking algorithms, and, by extension, admitted classes. We seek to address this by \textbf{quantifying the impact of different admission policies on the applications prioritized for review}. We show that removing race data from a previously developed applicant ranking algorithm reduces the diversity of the top-ranked pool of applicants without meaningfully increasing the academic merit of that pool. We \textit{contextualize this impact} by showing that excluding data on applicant race has a greater impact than excluding other potentially informative variables like intended majors. Finally, we \textit{measure the impact of policy change on individuals} by comparing the arbitrariness in applicant rank attributable to policy change to the arbitrariness attributable to randomness (i.e., how much an applicant's rank changes across models that use the same policy but are trained on bootstrapped samples from the same dataset). We find that any given policy has a high degree of arbitrariness (i.e. at most 9\% of applicants are consistently ranked in the top 20\%), and that removing race data from the ranking algorithm increases arbitrariness in outcomes for most applicants.\looseness=-1 
\end{abstract}

\ccsdesc[500]{Applied computing~Law}
\ccsdesc[500]{Social and professional topics~User characteristics}
\ccsdesc[500]{Computing methodologies~Machine learning}

\keywords{college admissions, affirmative action, machine learning, ranking system, fairness, arbitrariness}

\maketitle

\section{Introduction}\label{s-intro}

At selective American colleges, admissions is a high-stakes decision-making process in which reviewers must sort through a large pool of applicants in order to admit a first-year class that displays both \textit{academic merit} and \textit{diversity} \cite{princeton1977admissions, MITadmissions}. Many such institutions currently rely on \textit{holistic admissions}, which purport to evaluate the `whole' student -- not only through metrics like test scores and GPA, but also through more subjective factors like background and life experiences \cite{bastedo2018what, collegeboard2018, stevens_2007_creating}. Holistic admissions is a labor-intensive process that requires human review of a complex array of information (grades, activities, essays, letters of recommendation, etc.) for a large volume of applications (typically numbering in the tens of thousands) in a small amount of time (around three months).\footnote{See Appendix \ref{a-1} for details of application volumes and timelines at selective American colleges.}
Because the number of applications has been steadily growing year-over-year, this process has become increasingly challenging to scale \cite{NCES2023applications}. A critical question for admissions offices is therefore how to prioritize applications for review in order to most effectively make use of the limited time that reviewers have. In the past, standardized test scores (SAT, ACT, etc.) have provided a common method for ranking or at least grouping applicants to organize the review process \cite{lee2023evaluating}. However, test-optional policies enacted during the COVID-19 pandemic have limited colleges' access to those standardized metrics, forcing admissions offices to again grapple with the question of how to best organize applicants \cite{nyt2024misguided}. Any one piece of information contained in an application may be too coarse, inconsistently measured, or insufficiently indicative of potential success to impose a ranking on the full applicant pool. Therefore, researchers are increasingly exploring machine learning (ML) as an approach to identify complex relationships between the information contained in historic applications and their corresponding admissions decisions in order to prioritize applicants for review \cite{lee2023evaluating, waters_grade_2014, ragab2012hrspca, sridhar2020university, vaghela2015students}.\looseness=-1

This raises an important question: what information should these algorithms take into consideration when prioritizing applicants for review? Prior research efforts have taken inspiration from the ethos of holistic admissions and included all data available in an application, including not only grades, essays, and test scores, but also details of an applicant's background including their race and ethnicity \cite{lee2023evaluating}. Until recently, this approach was justifiable under the policy of \textit{affirmative action}, in which applicants belonging to historically marginalized groups were given special consideration in the admissions process. Selective colleges argued that affirmative action benefited not only those applicants, but also the student body as a whole, pointing out that ``a great deal of learning [...] occurs through interactions among students [...] who are able, directly or indirectly, to learn from their differences'' \cite{princeton1977admissions}. However, in June 2023, affirmative action in college admissions was effectively abolished by the U.S. Supreme Court, which ruled in Students for Fair Admissions v. Harvard that it constituted a form of racial discrimination and was thus unconstitutional \cite{sc2023students}. As a result, elements of the admissions process that previously considered race and ethnicity must be updated to exclude those variables. There is currently a poor understanding of how these mandated changes will shape how applicants are prioritized for review, and, by extension, who is admitted.\looseness=-1

\subsection{Research Questions and Contributions}\label{s-1.1}

In this work, we rely on four years of admissions data from a selective American higher education institution to explore how changes in admission policy -- especially the end of affirmative action -- impact admission processes. In particular, we address the following research questions:
\begin{itemize}
\item \textbf{RQ1}: How does a change in admission policy impact a college's \textit{overall class}?
\item \textbf{RQ2}: How does a change in admission policy impact \textit{individual applicants}?
\end{itemize}

We cannot observe actual admissions outcomes under different policies. As a proxy, we therefore rely on applicant ranking algorithms, which can be used to determine the order in which applicants are reviewed. We argue that the rankings produced by these algorithms under different policies are likely to meaningfully impact admissions. Even if we assume that the order in which an applicant is reviewed does not impact how they are rated -- that is, an applicant has roughly the same chance of being considered a `good' candidate whether they are reviewed first, last, or somewhere in the middle -- review order can still impact outcomes. There are constraints on the size of a college's first-year class: there are typically vastly more `good' candidates than admission slots, meaning that only a subset can be accepted. Based on conversations with the admissions team at the case institution, we model the admissions officers as adding applicants to the pool of accepted students as they review their applications and deem them `good' candidates. This means that the class may be full before later-reviewed applicants can be added, even if they are deemed equally `good' by admissions officers. Therefore, we define impact according to \textit{the order in which applicants are prioritized by a ranking algorithm.} Again based on conversations with the case admissions office, we pay special attention to the set of applicants that are designated as part of a `top' pool of applicants. We seek to not only \textit{measure} but also \textit{contextualize} this impact. In addition to measuring how the new ban on consideration of race is likely to change the applicants that are placed in the top pool by ranking algorithms, we also compare these results to those of other hypothetical policy changes. Finally, incorporating recent scholarship on \textit{model multiplicity} \cite{black_model_2022} and \textit{arbitrariness} of predictive model decisions \cite{cooper2024arbitrariness}, we go beyond considering the impact of policy changes on the top-ranked pool in aggregate and examine the likely impact on individual applicants. In particular, we compare the change in each applicant's rank under different policies to random changes in their rank, induced via bootstrapping, within the same policy. \looseness=-1

Together, this approach allows us to (1) predict the likely impact of race-unaware admissions on the ability of colleges to admit a first-year class that displays both academic merit and diversity; (2) contextualize this impact by comparing changes in top-ranked applicants under a race-unaware policy to changes under other hypothetical policies; and (3) provide a template for going beyond group fairness assessments of college admissions to understand the relative impact of policy changes on individual applicants. Ultimately, we find that: (1) \textbf{race-unaware policies do not meaningfully improve the academic merit of the top-ranked pool even as they significantly decrease diversity}, countering narratives about the costs of affirmative action. (2) \textbf{Excluding race data has a greater impact on diversity than other hypothetical policy changes}, such as excluding data on intended majors. At the individual level, we find that (3) \textbf{any given policy has a high degree of arbitrariness} (i.e. at most 9\% of applicants are consistently ranked in the top 20\% by any policy). We also find that (4) \textbf{individual applicant outcomes exhibit about 70\% more arbitrariness across policies than within a given policy}. Further, (5) \textbf{under a race-unaware applicant ranking algorithm, arbitrariness resulting from inherent randomness increases relative to the baseline for most applicants} (including, on average, Asian, Hispanic, white, and multi-racial applicants).\looseness=-1
\section{Background and Related Work}\label{s-related-work}
In this section, we outline the goals and challenges of modern-day admissions policies at selective American colleges and how ML has been applied in support of those goals (\S\ref{s-2.1}). We also provide an overview of how researchers and activists have measured the impact of admissions processes and policies on school applicants, highlighting gaps in prior assessment methods that we seek to address with this work (\S\ref{s-2.2}). We note at the outset that we focus here on \textit{selective, American} colleges and universities, as these are the institutions whose admissions processes are most likely to be impacted by the Supreme Court's recent ban on affirmative action. Colleges outside of the U.S. are not subject to the ruling; non-selective colleges by definition admit the majority of their applicants and as such typically do not consider race in admissions \cite{brookings2023admissions, nyt_selective}.\looseness=-1 

\subsection{Selective College Admissions: Goals and Challenges}\label{s-2.1}

\paragraph{Admitting applicants with academic merit}
Perhaps the most important goal of college admissions is identifying students with academic merit, typically defined as those who are predicted to succeed academically if admitted to the institution \cite{CMUadmissions, ColumbiaAdmissions, StanfordAdmissions}. As the number of college applications submitted reaches historic highs, this process has become increasingly labor-intensive. According to the National Center for Education Statistics, college applications increased by 36\% between 2014 and 2022, from 9.6 to 13.1 million \cite{NCES2023applications}. Selective colleges must now review tens of thousands of applications in order to fill just a few thousand slots. At the same time, the data available to them to make these decisions is changing. The suspension of standardized tests like the SATs during the COVID-19 pandemic led many colleges to go `test optional.' Among the approximately one thousand colleges that rely on the Common App, a platform through which millions of college applications in the U.S. are submitted annually, 55\% required standardized test scores in 2019, but in 2023, this number has plummeted to 4\% \cite{common2024deadline}. Applicants are embracing these test-optional policies: the Common App reports that 76\% of applicants submitted test scores in 2019, compared with just 45\% in 2023 \cite{common2024deadline}. As a result, colleges no longer have reliable access to a key piece of nationally standardized information that has been shown to be predictive of student success \cite{chetty_diversifying_2023, friedman_2024_standardized}.\looseness=-1 

\paragraph{Admitting a diverse class} Another goal of college admissions is diversity. For most of American history, access to higher education was largely restricted to those who were white, male, and Protestant. Slavery, coupled with anti-literacy legislation in the South and laws barring Black students from public schools in many Northern states, restricted access to education for most Black Americans before the Civil War \cite{ncantiliteracy, indianalaw}. Following the abolition of slavery, this discrimination persisted in the form of legally codified segregation that provided Black Americans with lower-quality education. Racial minorities, religious minorities, and women who did apply for higher education were often rejected due to outright bans or quotas restricting their admission; those who were admitted faced segregation and other forms of discrimination \cite{harvardlegacy}. In the 1960s, spurred by nationwide civil rights protests, selective colleges began adopting affirmative action policies to admit Black and other minority students who, as a direct result of this systemic discrimination, did not have comparable grades and test scores to their white peers \cite{stulberg2014origins}. Colleges argued that affirmative action was beneficial not only to historically disadvantaged students, but to the institution as a whole due to the ``educational benefit that flows from student body diversity'' \cite{sc2003grutter}. Nevertheless, decades of legal challenges limiting affirmative action \cite{sc1978bakke, sc2003gratz} culminated in the Supreme Court ruling in 2023 that race-conscious admissions ``violate the Equal Protection Clause of the 14th Amendment'' \cite{sc2023students}, effectively banning the use of race data in college admissions. We refer to this as the \textit{SFFA\footnote{After Students for Fair Admission Inc., (SFFA), who brought the suit leading to the Supreme Court ruling.} policy change} throughout.\looseness=-1

\paragraph{Machine learning in admissions processes}
In the face of these challenges, colleges are increasingly turning to ML to aid their admissions processes \cite{mcconvey_human-centered_2023}. One common, and potentially fraught, use case is to rank applicants using ML (typically by generating scores corresponding to applicants' predicted chance of admission) and provide the ranking to human reviewers in order to speed up or scale the admissions process \cite{shao2022combinatorial, staudaher2020prediction, waters_grade_2014}. For example, GRADE, a tool used for graduate admissions at the University of Texas at Austin, was developed because ``the number of applications [had] become too large to manage with a traditional review process'' \cite{waters_grade_2014}. At UT Austin, reviewers were asked to `validate' applicants to whom GRADE had given very high or low scores and focus most of their time and energy on reviewing applicants about whom GRADE was unsure \cite{waters_grade_2014}. GRADE was used for years before it was abandoned in 2020 due to widespread concerns that it was reinforcing historical biases in admissions \cite{IEDGRADE}. Although GRADE's creators argued that it admitted students based purely on academic merit \cite{waters_grade_2014}, critics pointed to its reliance on variables like the `eliteness' of applicants' undergraduate institutions, reinforcement of potentially biased historic decisions, and lack of a bias assessment during the tool's development and deployment to successfully argue for its discontinuation \cite{IEDGRADE}.\looseness=-1

However, the use of non-ML based applicant ranking and selection approaches can also be controversial. Even deciding whether to rely on standardized test scores, for example, entails a complex tradeoff \cite{garg_standardized_2021}: although research has shown that test scores display racial and socioeconomic gaps that may not be reflective of merit \cite{reardon_widening_2011}, there is also evidence that those same scores improve the ability of colleges to identify under-represented applicants with high academic merit \cite{UCreport, dartmouthSAT}, and that they may encode less bias than other, more subjective, application materials \cite{dutt_gender_2016}. In this context, multiple researchers have explored the possibility of using \textit{fairness-aware} ML to improve diversity in admissions. \citet{alvero_ai_2020}, for example, found that even simple natural language processing (NLP) models were able to distinguish between college application essays written by students of different income levels and genders. \citet{lee_augmenting_2023} subsequently quantified the impact of using this information to improve diversity in an admissions decision support algorithm, finding that essay data helped improve gender diversity, but did not have a significant impact on racial diversity. In a similar study, \citet{lee2023evaluating} took a fairness-aware approach to build an ML-based applicant ranking algorithm to replace standardized test scores, explicitly considering demographic features like race along with other holistic variables in order to increase an institution's ability to identify a diverse set of students with high academic merit. We contribute to this prior work by taking a fairness-aware approach to explore how applicant ranking algorithms based on a broad set of features are likely to change under the SFFA policy change. \looseness=-1 

In addition, we note that there has been significant work done to develop fair ranking algorithms more generally (i.e. not specifically within the domain of college admissions) \cite{zehlike_2017_fair, asudeh_2019_designing, zehlike_2022_fairness, zehlike_2022_fairnessII, singh_2019_policy, singh_2018_fairness, celis2018ranking, patro2022fair}. However, the applicant ranking algorithms of which we are aware do not tend to incorporate these fair ranking approaches \cite{lee2023evaluating, waters_grade_2014}; moreover, to the extent that fair ranking mechanisms require access to demographic data, they will not be feasible in the college admissions domain in the future due to the SFFA policy change. As a result, we do not explore fair ranking algorithms in this work, which is intended not to \textit{recommend} approaches for building applicant ranking algorithms, but rather to \textit{predict} how the SFFA policy change will impact already-existing processes. We leave the development of such approaches, that are compatible with the legal environment, for future work.\looseness=-1

\subsection{Measuring the Impact of Admission Processes and Policies on College Applicants}\label{s-2.2}

\paragraph{Impact on groups of applicants}
Given the importance of education and the fraught nature of admissions, many researchers have sought to measure the impact of admissions processes on applicants. Many of these assessments focus on \textit{demographic fairness}, which is typically defined as disproportionate admission rates across demographic groups including race, gender, and socioeconomic status. In two notable and recent studies, both \citet{grossman_disparate_2023} and \citet{chetty_diversifying_2023} conducted large-scale analyses on application data to quantify admissions disparities across demographic groups and identify features driving those disparities. \citet{grossman_disparate_2023} focused on race, finding that Asian students were significantly less likely to be admitted to selective American colleges than white students with comparable test scores, grades, and extracurriculars, and that this disparity was partially (but not entirely) driven by legacy admissions and geographic considerations. \citet{chetty_diversifying_2023} focused on socioeconomic status, finding that students from families with incomes in the top 1\% of the U.S. are more likely to be admitted to Ivy-Plus colleges, and that this disparity is mostly driven by factors like recruited athlete status, legacy status, and `non-academic' (e.g. extracurricular) ratings. Both papers also explore admissions policies that could alleviate these disparities by reconsidering how achievement indicators and sociodemographic attributes are considered in the process \cite{grossman_disparate_2023, chetty_diversifying_2023}.\looseness=-1

Other researchers have taken more qualitative approaches to examine fairness as well as \textit{transparency} and \textit{trust} in admissions processes. For example, through field observation and a series of interviews at the University of Oxford, \citet{zimdars_fairness_2010} found evidence that unconscious bias by reviewers led to disproportionately high admission rates for applicants who were white, male, and members of the professional class. \citet{marian_algorithmic_2023} crowdsourced data on school assignments in New York City to identify the factors that impacted students' outcomes in order to increase transparency in the process. Similarly, \citet{robertson_modeling_2021} explored school assignment algorithms in San Francisco through the lens of value-sensitive design in order to understand why changes intended to increase diversity had in practice exacerbated school segregation.\looseness=-1 

\paragraph{Our contribution: impact on applicants as individuals} 
Missing from these assessments is an examination of \textit{arbitrariness} in college admissions. Prior work on \textit{model multiplicity} has shown that predictive tasks can be satisfied by multiple models that are equally accurate (e.g., at identifying students with academic merit) but differ in terms of their individual predictions (e.g., whether a specific applicant is considered a top applicant) \cite{black_model_2022, watsondaniels_multitarget_2023}. This means that any number of seemingly minor decisions made by a college admissions office -- down to something as simple as how to sample training data -- could impact an individual applicant's outcome even if it does not majorly affect the ability of the college to identify students with academic merit. In fact, \citet{cooper2024arbitrariness} show that training otherwise identical models on bootstrapped samples of the same dataset can result in predictions for individuals that are essentially arbitrary: half the time, individuals are predicted to belong to one binary class, and half the time to the other. Similar, seemingly arbitrary, choices, such as how to pre-process variables, have also been shown to cause major changes in model outputs \cite{huynh_2024_mitigating}. There is significant evidence that there are more applicants with high academic merit in a typical pool than there are open seats in a college's first-year class. In particular, a large proportion of applicants with perfect or near-perfect test scores are rejected by selective colleges \cite{grossman_disparate_2023}; additionally, each year, selective colleges offer positions on their waitlist to hundreds if not thousands of ``[s]tudents who met admission requirements but whose final admission was contingent on space availability'' \cite{CDS}.\footnote{See Appendix \ref{a-1} for details of waitlist sizes for selective American colleges.} Given this, it is likely that, if arbitrariness resulting from minor modeling choices impacts how an applicant is prioritized for review by an ML model, then it could also have a downstream impact on their final admissions outcome.\footnote{This follows from the assumption we describe in \S\ref{s-1.1}: that admissions officers add applicants to the pool of accepted students as they review their applications and deem them `good' candidates.} We propose that in order to fully understand the impact of admission processes and policies on college applicants, researchers must consider not only fairness and diversity in group outcomes, but also fairness and arbitrariness in individual outcomes.\looseness=-1

\section{Method}\label{s-method}

We seek to quantify and contextualize the impact that the SFFA policy change is likely to have on which applicants are prioritized for review (i.e., ranked in the top category by an ML algorithm). In this section, we describe the data available to us and provide a brief overview of the first-year undergraduate admissions process at our case institution (\S\ref{s-3.1}); we also describe our baseline ranking models (\S\ref{s-3.2}) and simulated policy changes (\S\ref{s-3.3}). In addition, we outline our approach to measuring and contextualizing expected changes in the predictive power and diversity of overall rankings (\S\ref{s-3.4}) and individual applicant outcomes (\S\ref{s-3.5}) due to the SFFA policy change.\looseness=-1

\subsection{Background and Ranking Algorithm}\label{s-3.1}
\paragraph{Case institution.} Our case institution is a highly selective, engineering-focused American university that requires first-year undergraduate applications to be submitted through the Common App. Through the Common App, the case institution collects applicants' standardized test scores, high school grades and coursework, extracurricular activities, family and demographic background (including race and ethnicity, citizenship, and parental education), essays, and letters of recommendation. The case institution has been test-optional since the 2020-2021 admissions cycle. The data that we use for this analysis spans all Regular Decision applicants from the 2019-2020 admissions cycle to the 2022-2023 admissions cycle (four years of applications, or 59,833 in total). During the 2019-2020 admissions cycle (prior to the test-optional policy), 92\% of applicants submitted either SAT or ACT scores;\footnote{We assume that the remaining 8\% of students submitted their scores late, meaning we do not have access to that data.} in the years since, an average of 67\% of applicants have submitted either SAT or ACT scores annually. At the case institution, each application is reviewed twice: first by a seasonal reader, and then by a member of the staff of the admissions office. Applications were historically prioritized for review based on standardized test scores; following the establishment of the test-optional policy, the admissions office has explored using ML-generated scores instead.\looseness=-1 

\paragraph{Preprocessing.}
To understand how the SFFA policy change is likely to impact \textit{already existing} applicant ranking algorithms, we followed similar data acquisition and preprocessing steps to those already implemented by the case institution. We included all data that was common across all four years of applications, except for personally identifiable information (name, date of birth, contact information, etc.) and data that is not entered directly into the Common App form but is instead provided as a file upload (transcript, letters of recommendation, etc.). Ultimately, this left us with 302 raw features; as part of preprocessing (following a similar approach to that outlined in \cite{lee2023evaluating}), we also applied a one-hot encoding to categorical features, recoded categories that occur in fewer than 1\% of observations as `RARE', imputed missing values and added indicator variables indicating that a numeric feature was missing (categorical variables were directly imputed as `MISSING'), and constructed TF-IDF unigrams and bigrams for text features. We split our data into train and test sets based on year: we used the 2019-2020, 2020-2021, and 2021-2022 admissions cycles as training data and the 2022-2023 admissions cycle as test data. This mimics the real-world scenario of training a model on all available historical data, and also allows us to use results for a complete applicant pool as test data. Summary statistics describing our training and test data sets are presented in Table \ref{table-summary-stats}.\looseness=-1

\begin{table*}[t]
\centering
\caption{Summary statistics for our training and test data. The training set includes three years of data from 2019-2020, 2020-2021, and 2021-2022 admissions cycles, and the test set includes the 2022-2023 cycle.}
\label{table-summary-stats}
\begin{tblr}{
  row{1} = {c},
  cell{2}{2} = {c},
  cell{2}{3} = {c},
  cell{2}{4} = {c},
  cell{2}{5} = {c},
  cell{2}{6} = {c},
  cell{2}{7} = {c},
  cell{2}{8} = {c},
  cell{3}{2} = {c},
  cell{3}{3} = {c},
  cell{3}{4} = {c},
  cell{3}{5} = {c},
  cell{3}{6} = {c},
  cell{3}{7} = {c},
  cell{3}{8} = {c},
  hline{1,4} = {-}{0.08em},
  hline{2} = {-}{0.05em},
}
 & \textbf{\# applicants} & \textbf{\% accepted} & \textbf{\% waitlist} & \textbf{\% URM} & \textbf{\% female} & \textbf{\% first-generation} & \textbf{\% low SES}\\
\textbf{Train} & 44,293  & 5.7\% & 14.4\% & 17.3\% & 31.1\% & 15.9\% & 25.7\%\\
\textbf{Test} & 15,540  & 5.0\% & 16.1\% & 16.3\% & 32.1\% & 19.5\% & 32.3\%
\end{tblr}
\end{table*}

\paragraph{Modeling.} Finally, we formulate a probabilistic classification task where applicants in the training data are categorized as `admitted' if they are either outright or conditionally admitted and `not admitted' otherwise.\footnote{However, we note that the selection of target variable is a non-trivial decision \cite{passi_problem_2019} and therefore provide an analysis of the robustness of our approach with alternative target variable selection in Appendix \ref{a-3}.} We use a Gradient-Boosted Decision Tree as our classifier. Rather than predict a binary `admitted'/`not admitted' outcome for each applicant in the test data, we instead segmented applicants into deciles based on their predicted probability of admission. This follows the approach outlined in \citet{lee2023evaluating} and also aligns with how admissions offices are likely to use applicant ranking algorithms in practice: as a way of prioritizing applications for review, and not as a way of directly admitting applicants. Decile 1 contains the 10\% with the lowest predicted probability of admission; Decile 10 contains the 10\% with the highest probability of admission. Following \citet{lee2023evaluating}, we further segment these deciles into `top,' `middle,' and `bottom' pools. Based on approaches previously explored by the case institution, we use Decile 9 and Decile 10 as our top pool: this is the set of applicants with the highest expected number of admissible students according to the ranking algorithm, and is thus the set that the admissions office wants to review first. In order to ensure that our results are not brittle to our specific choice of cutoff for assigning the top pool, we verify via a robustness assessment that our findings are consistent across decile cutoffs. The results of that analysis are shown in Appendix \ref{a-2}.\looseness=-1

\subsection{Defining Baselines}\label{s-3.2}
In order to measure the impact of the SFFA policy change, we first need to establish a baseline to compare against. Here we rely on two baselines. First, we train an \textit{ML baseline} model that uses every variable available from the processed Common App data (as described above) to predict an applicant's likelihood of being accepted. Following \citet{lee2023evaluating}, we believe that this model is a reasonable representation of how college admissions offices might incorporate ML into their processes.\looseness=-1

We also define a \textit{naive baseline} model that ranks applicants based on their prior math instruction and their standardized test scores. \citet{lee2023evaluating} suggest that relying on standardized test scores represents an applicant ranking method that was commonly used pre-COVID. Because our test set contains data from the 2022-2023 admissions cycle (i.e. after test-optional policies were put in place at our case institution), we are missing standardized test scores for 32\% of the applicant pool. Therefore, we supplement our baseline ranking with data on applicants' highest level of math taken, which is both salient to our (engineering-focused) case institution and available for the entire pool of applicants. We first rank applicants according to their highest math course taken.\footnote{We note that, by ranking students according to their highest math class taken, the baseline model likely penalizes URM students for factors beyond their control. The U.S. Department of Education's Office for Civil Rights has reported that just 38\% of public high schools with high ($\geq$75\%) Black and Hispanic enrollment offer calculus, compared with 50\% of public high schools nationally \cite{DOEmath}. We do not suggest this model be used to rank applicants in practice and only put it forward as a baseline.} Next, we convert applicants' reported SAT\footnote{\url{https://research.collegeboard.org/reports/sat-suite/understanding-scores/sat}} and ACT\footnote{\url{https://www.act.org/content/act/en/products-and-services/the-act/scores/national-ranks.html}} scores to percentiles in order to make them directly comparable with one another. Then, within the band of `highest math taken,' we rank applicants according to the higher of their two percentiles (applicants who did not report either score are ranked last within their band\footnote{How to rank students who do not report a test is a non-trivial choice due to informational differences, fairness concerns to both those who report and do not report scores, and concerns about strategic reporting behavior \cite{liu2021test}. However, because we consider test scores only within a math course band, the effect of this choice is relatively small.}). Finally, we again select the top pool of applicants where we assume an admissions office would focus the majority of their attention. Mirroring the ML approach described above, we define this as the set of applicants with the top 20\% of baseline scores.\footnote{However, we note that because this metric is coarser than our predictive model, there are a large number of ties among top-scoring applicants. Taking ties into account, 21\% of applicants share the top 20\% of scores.}

\subsection{Simulating Policy Changes}\label{s-3.3}
To analyze the impact of the SFFA policy change and compare it to the impact of other hypothetical policy changes, we repeatedly omitted variables of interest from the ML baseline model. We modeled three different policy changes:
\begin{itemize}
\item \textbf{No race}: We removed 12 race-related features, including indicators for applicants' race and ethnicity and an indicator for whether an applicant belongs to a URM group. These represent the variables that must be excluded from the admissions process going forward as a result of the SFFA policy change. 
\item \textbf{No major}: We removed 1 feature describing applicants' intended major. This model is intended to contextualize the impact of excluding race from applicant ranking algorithms by allowing us to compare it to excluding another theoretically important feature that groups applicants but does not indicate membership in a historically disadvantaged group. Intended major is theoretically important because a much smaller percentage of applicants indicating a highly popular intended major, such as Computer Science, can be admitted given institutional constraints on major size.
\item \textbf{No uncontrollable features}: We removed 29 features that represent \textit{uncontrollable} elements of an application (compared to \textit{controllable} features, which are elements like major or test scores that an applicant can choose or change). The uncontrollable features we identified include all race-related features along with applicant sex, socioeconomic status, citizenship, family education, and type of school attended. This model is intended to contextualize the impact of excluding race from applicant ranking algorithms by allowing us to compare it to excluding \textit{all} uncontrollable features.
\end{itemize}

\subsection{Group Impact: Measuring Changes in Academic Merit and Diversity of the Top-Ranked Applicant Pool}\label{s-3.4}

\paragraph{Measuring academic merit of the top pool.} An important consideration for any applicant ranking algorithm is whether it is able to successfully identify applicants with high academic merit who should be prioritized for review by the admissions office. While the `academic merit' of an applicant is not directly measurable \cite{jacobs_measurement_2021},
we define two proxies. First, we rely on labels provided by the case institution and say that the academic merit of a top pool increases as the proportion of applicants who were actually admitted or waitlisted increases. We consider applicants who were not only admitted but also waitlisted because those applicants represent ``[s]tudents who met admission requirements but whose final admission was contingent on space availability'' \cite{CDS}, i.e. students with high academic merit who would be admitted if there was space for them. Second, in an attempt to isolate a measure of academic merit not dependent on historical decisions, we also use the average percentile ranking of test scores submitted by applicants in the top pool. To measure whether policy changes result in statistically significant differences in the ability of ranking algorithms to identify applicants who were actually admitted or waitlisted, we conduct a binomial test comparing the share of those applicants identified in the top pool by the ML baseline model to the share identified by all other models. Similarly, to measure whether policy changes result in statistically significant differences in the ability of ranking algorithms to identify applicants with high test scores, we used the Mann-Whitney $U$ test to compare across the ML baseline and all other models. To determine statistical significance, we use a standard $p = 0.05$ threshold and apply the Bonferroni correction to account for multiple comparisons.\looseness=-1

\paragraph{Measuring diversity of the top pool.} We define the diversity of our top-ranked pool according to several factors. First, we look at the breakdown of applicants according to their self-identified race and ethnicity (based on the categories available in the Common App). We also consider the share of applicants who are under-represented minorities (URM).\footnote{We define URM in the same way as the Common App, which classifies ``Black or African American, Latinx, American Indian or Alaska Native, or Native Hawaiian or Other Pacific Islander'' applicants as URM applicants \cite{common2024deadline}.} Finally, we consider socioeconomic factors, including the share of applicants who identify as being the first in their family to attend college (first-gen), and the share of applicants with low family incomes (low-income).\footnote{We do not have direct access to applicants' family incomes status, so we use whether an applicant received an application fee waiver as a proxy; see: \url{https://appsupport.commonapp.org/applicantsupport/s/article/What-do-I-need-to-know-about-the-Common-App-fee-waiver}} We used binomial tests as described above to determine whether any policy changes result in statistically significant changes to the diversity of the top-ranked pool of applicants as compared to the ML baseline.\looseness=-1

It is important to note that the above is a narrow definition of diversity. URM status -- and for that matter, racial categorization as defined by the Common App -- is controversial and cannot fully represent an applicant's racial identity \cite{andrus_demographic-reliant_2022}; further, an applicant's contribution to a diverse campus cannot be reduced to their race/ethnicity and socioeconomic status. However, we choose to focus on these factors as we believe that URM and FGLI (first-generation and low-income) status are salient in the context of the SFFA policy change: opponents of the change worry that it will result in a decline in URM enrollment in selective American colleges specifically, and some have suggested mitigating this impact through an increased focus on socioeconomic diversity (i.e. FGLI status) in admissions instead \cite{brookings_end, nyt_affirmative}. \looseness=-1

\subsection{Individual Impact: Measuring Changes in Applicant Outcomes}\label{s-3.5}

\paragraph{Measuring arbitrariness across models.} For an individual applicant, arbitrariness across different ranking algorithms can be measured based on how consistently the applicant is placed in the top pool, or not placed in the top pool. To quantify this, we adapted the metric of \textit{self-consistency}, defined by \citet{cooper2024arbitrariness} as ``the probability that two models produced by the same learning process on different n-sized training datasets agree on their predictions for the same test instance'' \cite{cooper2024arbitrariness}. We present a modified version of this metric. Rather than consider only variation introduced by different training datasets, we also examine variation introduced by policy change (i.e. different learning processes) for models trained on the same dataset. For a given applicant, self-consistency is defined as:

\begin{equation}
   \textsc{sc} =  1 - \frac{2M_0M_1}{M(M-1)}\label{eq-self-consistency}
\end{equation}

\noindent where $M$ is the total number of models we examine, $M_1$ is the number of models in which an applicant is placed in the top pool and $M_0$ is the number of models in which an applicant is \textit{not} placed in the top pool. While self-consistency as defined by \citet{cooper2024arbitrariness} measures the consistency of decisions regardless of what those decisions are, we further distinguish between consistently being placed in the top pool and being consistently not placed in that pool throughout.

\paragraph{Comparing sources of arbitrariness}
To understand how much a policy change impacts an individual applicant's outcomes as compared to inherent randomness for a given policy, we compare self-consistency in applicants' outcomes \textit{across} different policies to their self-consistency \textit{within} bootstrapped instances of a single policy. To create a set of within-policy models $M_{within}$, we train one model on 1,000 bootstrapped samples of the training data. To create a set of across-policy models $M_{across}$, we sample 500 instances from the $M_{within}$ set of one model and 500 instances from the $M_{within}$ set of another model, creating a set of 1,000 modeling outcomes that represents two different policies. To compare the overall level of arbitrariness (the complement of self-consistency) \textit{across} vs.~\textit{within} policies, we calculate the \textit{arbitrariness ratio} \textsc{ar}:\looseness=-1 

\begin{equation}\label{eq-group-arbitrary}
    \textsc{ar} = \frac{1 - \bar{\textsc{sc}}_{across}}{1 - \bar{\textsc{sc}}_{within}}
\end{equation}

Intuitively, this is the ratio of the average level of arbitrariness across policies to the average level of arbitrariness within the ML baseline policy. The idea is that arbitrariness resulting from bootstrapping will be captured by both the $M_{within}$ and $M_{across}$ models, and any additional arbitrariness in the $M_{across}$ models will be attributable to policy change. We also test whether the overall level of arbitrariness across all applicants is statistically significantly different across vs.~within policies with a Wilcoxon signed-rank test, again with the standard $p = 0.05$ threshold. 

\paragraph{Predicting how arbitrariness will change as a result of the SFFA policy change}

Finally, we consider the fact that arbitrariness is not a fixed quantity: it can increase or decrease across ranking policies (as an intuitive example, the naive baseline has zero arbitrariness across repeated applications; a random ranking policy would have arbitrariness approaching 0.5). We therefore explore how individual arbitrariness resulting from random modeling choices changes if the ML baseline ranking algorithm is minimally modified to comply with the SFFA policy change (i.e. the `no race' model). We do this by conducting Wilcoxon signed-rank tests that compare the overall arbitrariness and arbitrariness of specific demographic groups between the ML baseline and `no race' models. As with our prior comparisons, we rely on a $p = 0.05$ threshold for significance and apply the Bonferroni correction to account for multiple comparisons.\looseness=-1
\section{Results}\label{s-results}
\subsection{Group Impact: Academic Merit and Diversity}\label{s-4.1}

\paragraph{Compliance with the SFFA policy change significantly reduces the diversity of top-ranked applicants} 
\begin{figure*}[h]
\centering
\subfloat[]{\includegraphics[height=4.5cm]{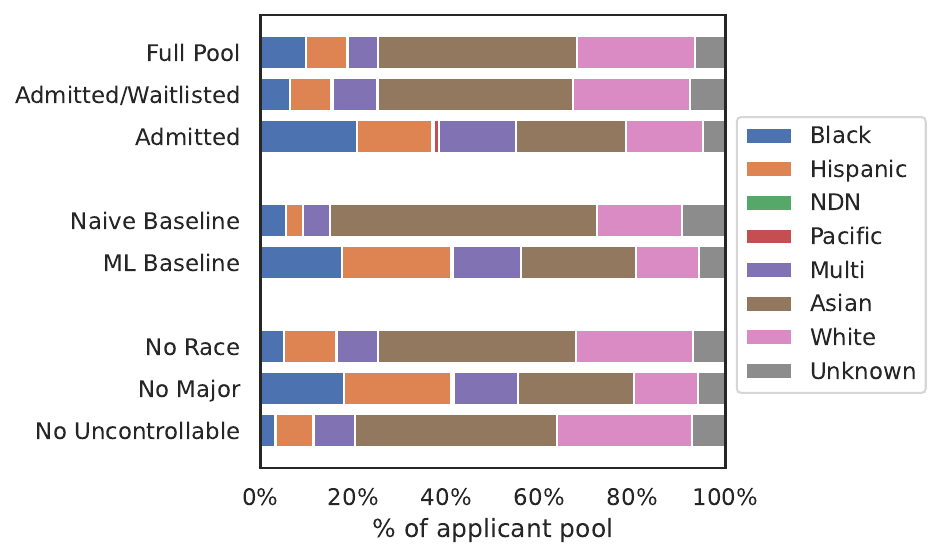}}
\hspace{0.05\textwidth}
\subfloat[]{\includegraphics[height=4.5cm]{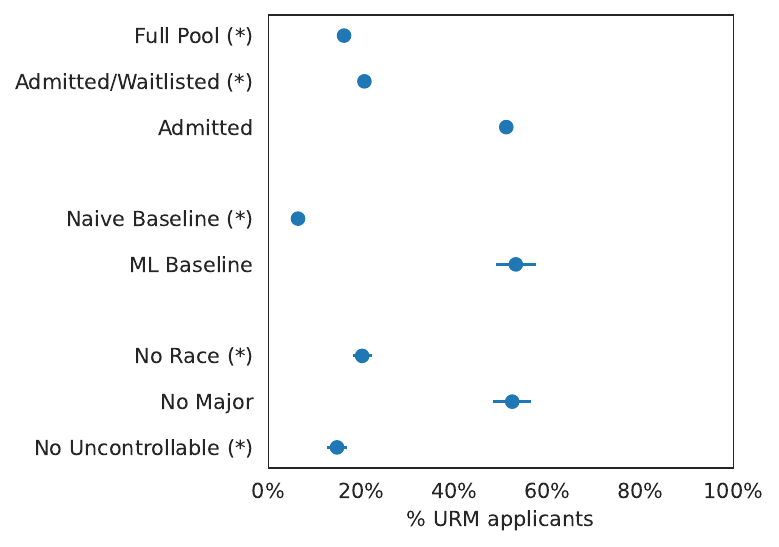}}
\Description{} 
\caption{Impact of policy changes on the \textbf{racial and ethnic diversity} of the top-rated group of applicants. Statistically significant differences in proportion of URM applicants present in the top-rated group of applicants compared to the ML baseline model are denoted with an asterisk. In Graph (b), 95\% confidence intervals for the ML models are shown based on results over 1,000 bootstraps.
\label{fig-group-diversity-race}}
\end{figure*}
\begin{figure*}[h]
\centering
\subfloat[]{\includegraphics[height=4.5cm]{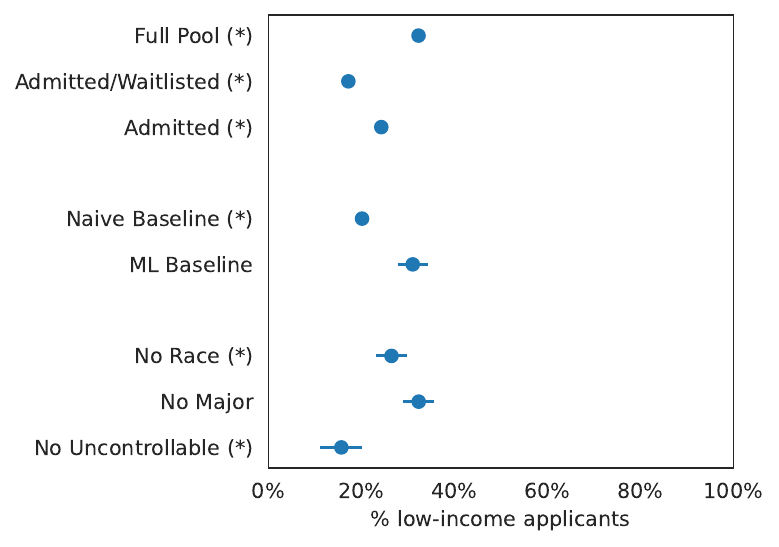}}
\hspace{0.05\textwidth}
\subfloat[]{\includegraphics[height=4.5cm]{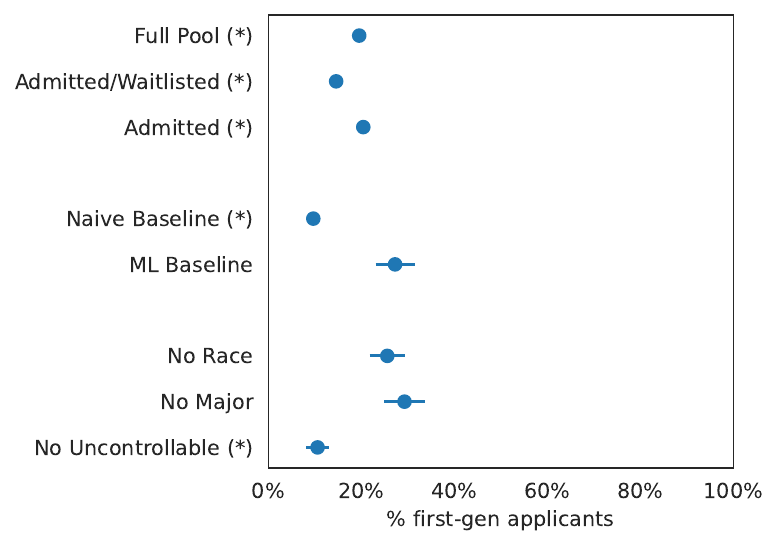}}
\Description{} 
\caption{Impact of policy changes on the \textbf{socioeconomic diversity} of the top-rated group of applicants. Statistically significant differences in proportion of first-gen and low SES applicants present in the top-rated group of applicants compared to the ML baseline model are denoted with an asterisk. 95\% confidence intervals for the ML models are shown based on results over 1,000 bootstraps.
\label{fig-group-diversity-ses}}
\end{figure*}

The ML baseline model, which uses all available Common App data to predict prior admissions decisions and which we argue represents a reasonable assumption of how admissions offices might previously have implemented applicant ranking algorithms, selects a top pool that is 53\% URM, as shown in Fig.~\ref{fig-group-diversity-race}(b). This over-represents URM applicants relative to the full applicant pool (16\%) and admitted or waitlisted group (21\%), but is in line with the share of URM applicants in the actually admitted group (51\%). When we exclude data related to applicant race and ethnicity from the ML baseline model, the share of URM applicants in the top-ranked pool falls to 20\%---a 62\% reduction compared to the ML baseline (this is statistically significant, with $p < 0.001$). Excluding other uncontrollable features, including data on applicants' gender and socioeconomic status, further exacerbates this reduction (15\%, $p < 0.001$). By contrast, excluding major preference from the applicant ranking algorithm results in a URM share of 52\%, which is not statistically significantly different from the ML baseline ($p = 0.39$).\looseness=-1 

Similar trends hold for socioeconomic diversity metrics, as shown in Fig.~\ref{fig-group-diversity-ses}. Excluding data related to race reduces the share of low-income applicants in the top pool by a small but statistically significant amount as compared to the ML baseline (from 31\% to 26\%, $p < 0.001$), but it does not significantly change the share of first-gen applicants (from 27\% to 26\%, $p = 0.04$\footnote{Not significant after accounting for multiple comparisons}). As with URM status, excluding all uncontrollable features further exacerbates the reduction in low-income (16\%, $p < 0.001$) and first-gen (11\%, $p < 0.001$) applicants in the top pool, but excluding applicants' intended major does not impact socioeconomic diversity as compared to the ML baseline (the share of low-income applicants increases to 32\% and the share of first-generation applicants increases to 29\%, but neither change is statistically significant ($p = 0.12$ and $p = 0.01$,\footnote{Not significant after accounting for multiple comparisons} respectively)).\looseness=-1

\paragraph{The reduction in diversity is not associated with a corresponding increase in academic merit of top-ranked applicants}

\begin{figure*}[tbh]
\centering
\subfloat[]{\includegraphics[height=4.5cm]{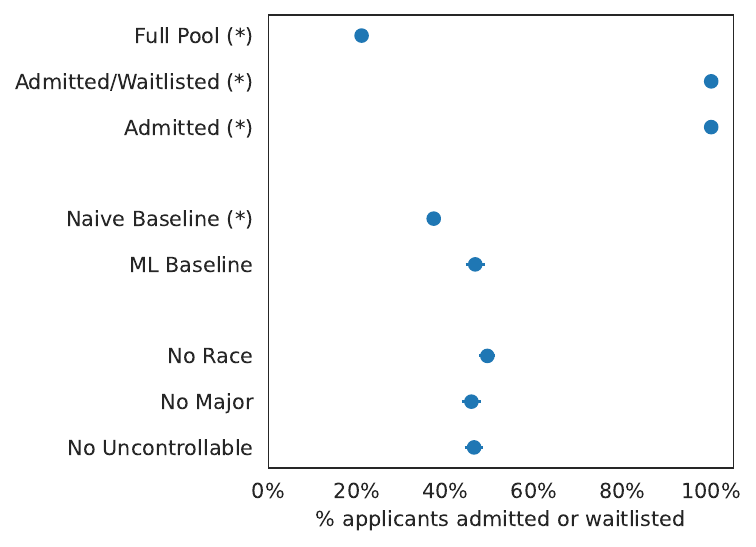}}
\hspace{0.05\textwidth}
\subfloat[]{\includegraphics[height=4.5cm]{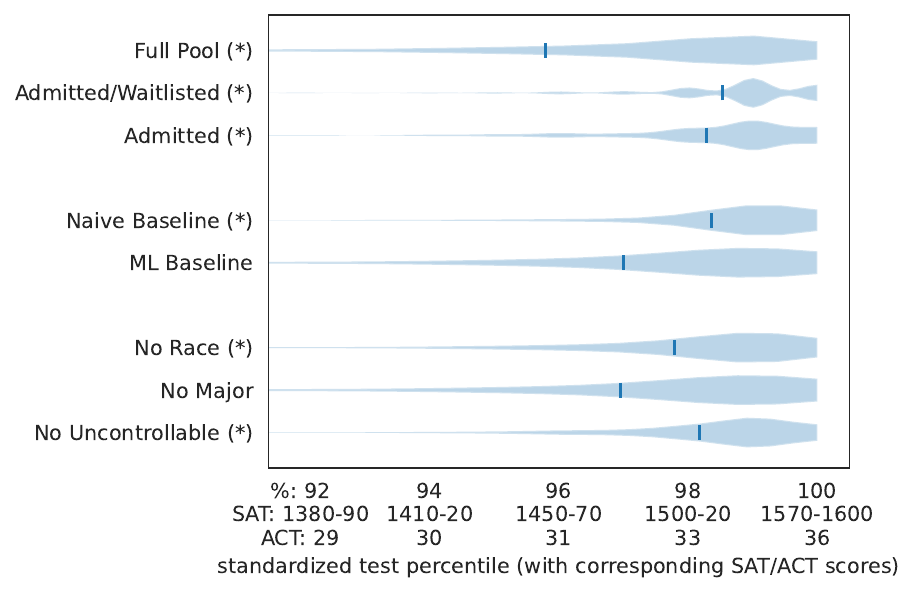}}
\Description{} 
\caption{Impact of policy changes on the \textbf{academic merit} of the top-rated group of applicants. Statistically significant differences in standardized test percentile and share of applicants actually admitted or waitlisted of the top-rated group of applicants compared with the ML baseline model are denoted with an asterisk. In Graph (a), 95\% confidence intervals for the ML models are shown based on results over 1,000 bootstraps. In Graph (b), the darker blue line represents the \textit{mean} standardized test percentile within the specified applicant pool. 
\label{fig-group-merit}}
\end{figure*}

Across all models, the academic merit of the top-ranked pool of applicants remains largely unchanged, as shown in Fig.~\ref{fig-group-merit}. The average standardized test percentile (among applicants who submitted standardized test scores) of admitted applicants was 98.3, compared to the ML baseline average of 97.0. Excluding race from the ML baseline model results in a statistically significant ($p < 0.001$) increase in the average standardized test percentile of the top-ranked students, to 97.8. In absolute terms, however, this is a small change: it is approximately the difference between a 1480 and a 1490 on the SAT. Excluding data on applicant major preference does not meaningfully change standardized test percentiles of the top-ranked pool ($p = 0.63$); excluding all uncontrollable features has a similar impact to excluding race alone ($p < 0.001$).\looseness=-1

In addition, other than the naive baseline ($p < 0.001$), which sorts students by highest math instruction and test scores, no model was statistically significantly better or worse than the ML baseline at identifying students who were actually admitted or waitlisted by the case institution. About half of the students included in the top pool are actually admitted or waitlisted across the ML baseline (47\%), No race (49\%), No major (46\%), and No uncontrollable features (46\%) models. \textbf{Overall, we predict that, if admissions ranking algorithms are minimally modified to comply with the SFFA policy change, they will prioritize a less diverse, but not more academically meritorious, pool of applicants for review.}\footnote{In Appendix \ref{a-2}, we show that this prediction is robust to the specific definition of `top' applicants.} \looseness=-1

\subsection{Individual Impact}\label{s-4.2}
\begin{figure*}[h]
\centering
\subfloat[]{\includegraphics[height=5cm]{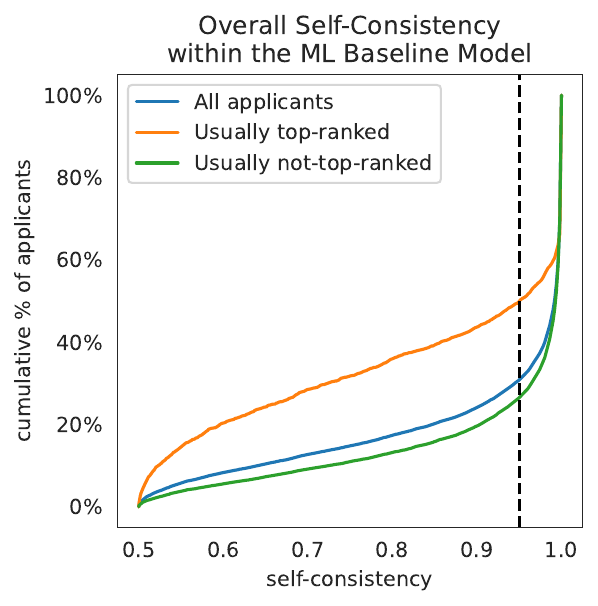}}
\hspace{0.05\textwidth}
\subfloat[]{\includegraphics[height=5cm]{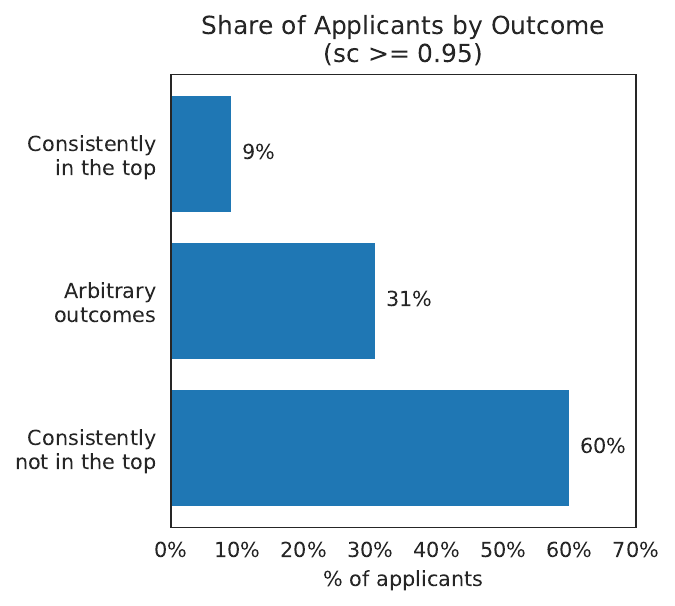}}
\Description{} 
\caption{Graph (a) shows the cumulative distribution (CDF) of self-consistency within 1,000 bootstraps of the ML baseline model for the applicant pool (blue line), only applicants who are usually top-ranked (ranked in the top by >50\% of bootstrapped models, orange line), and only applicants who are usually not top-ranked (ranked in the top by <=50\% of bootstrapped models, green line). The dashed black line corresponds to \textsc{sc} = 0.95. 
Graph (b) shows the level of arbitrariness if we define an applicant's outcomes to be consistent if their \textsc{sc} $\geq$ 0.95 (and their outcomes to be arbitrary if their \textsc{sc} $<$ 0.95): only 9\% of applicants are consistently ranked in the top, 60\% of applicants are consistently not ranked in the top, and 31\% of applicants have arbitrary outcomes.
\label{fig-within-MLB}}
\end{figure*}

\paragraph{Inherent randomness in the modeling process will lead to somewhat arbitrary outcomes, especially for top-ranked applicants} 

We calculated self-consistency for each applicant across 1,000 bootstraps of the ML baseline model, the cumulative distribution function of which is shown in Fig.~\ref{fig-within-MLB}(a). A self-consistency of 1 means that all (or none) of the bootstrapped models rank an applicant in the top pool, while a self-consistency of 0.5 means that exactly half of the models rank an applicant in the top pool. Most applicants have a relatively high self-consistency: across all applicants (blue curve), 31\% of applicants have the highest possible self-consistency, and 69\% of applicants have $\textsc{sc} \geq$ 0.95.\footnote{\textsc{sc} = 0.95 corresponds to agreement between 97.5\% of models, a very high level of agreement.} However, the ML baseline model more consistently identifies applicants who are \textit{not} included in the top pool (green curve) than applicants who \textit{are included} (orange curve). 

Fig.~\ref{fig-within-MLB}(b) provides deeper insight into the consistency of individual applicant outcomes at $\textsc{sc} \geq$ 0.95. It shows that over two-thirds of applicants (69\%) have consistent (or non-arbitrary) outcomes. However, just 9\% of these applicants are consistently ranked \textit{in the top pool}, with the remaining 60\% consistently ranked not in the top. Recall that the top pool consists of 20\% of the applicant pool: this means that in any given bootstrapped model, more than half of the top pool consists of applicants who have been added to that pool somewhat arbitrarily. While Fig.~\ref{fig-within-MLB}(c) shows that this is the case for \textsc{sc} $\geq$ 0.95, Fig.~\ref{fig-within-MLB}(a) shows that this effect holds across all self-consistency thresholds. Across the entire distribution, a larger proportion of usually-top-ranked applicants (applicants who are ranked in the top pool by >50\% of bootstrapped models) have lower self-consistencies than the usually-not-top-ranked applicants. Overall, we find that even within a single policy, randomness inherent to the modeling process has a major impact on who is included in the top pool.\looseness=-1

\paragraph{Within-policy arbitrariness increases under the SFFA policy}
\begin{figure*}[h]
\centering
\subfloat[]{\includegraphics[height=4.6cm]{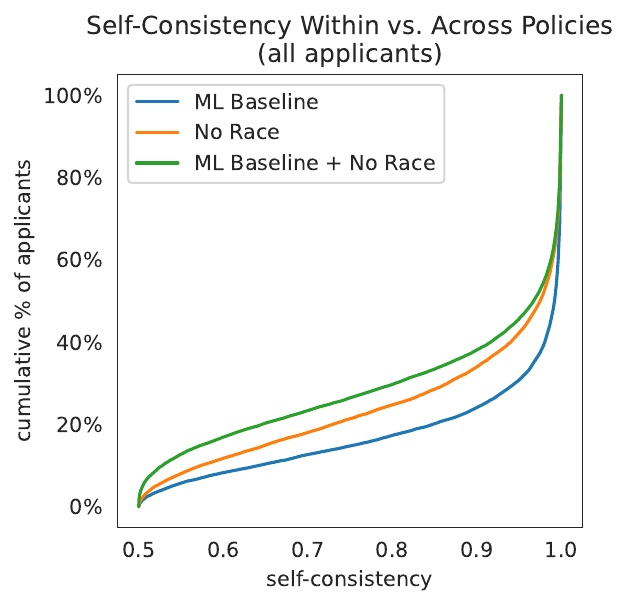}}
\hspace{0.01\textwidth}
\subfloat[]{\includegraphics[height=4.6cm]{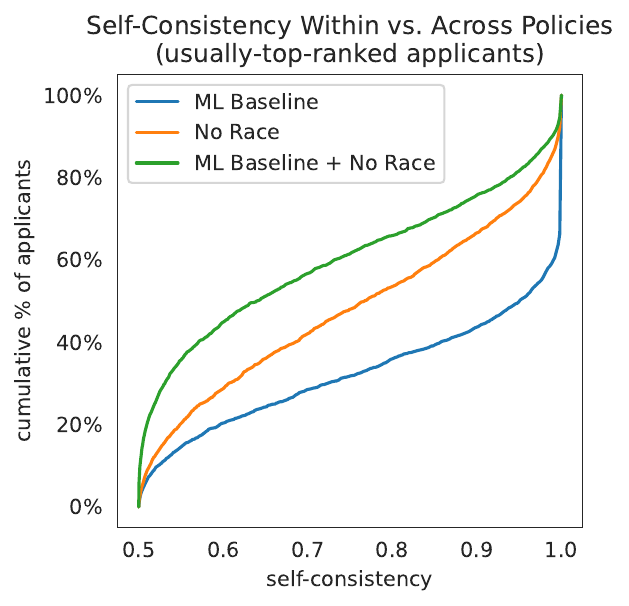}}
\hspace{0.01\textwidth}
\subfloat[]{\includegraphics[height=4.6cm]{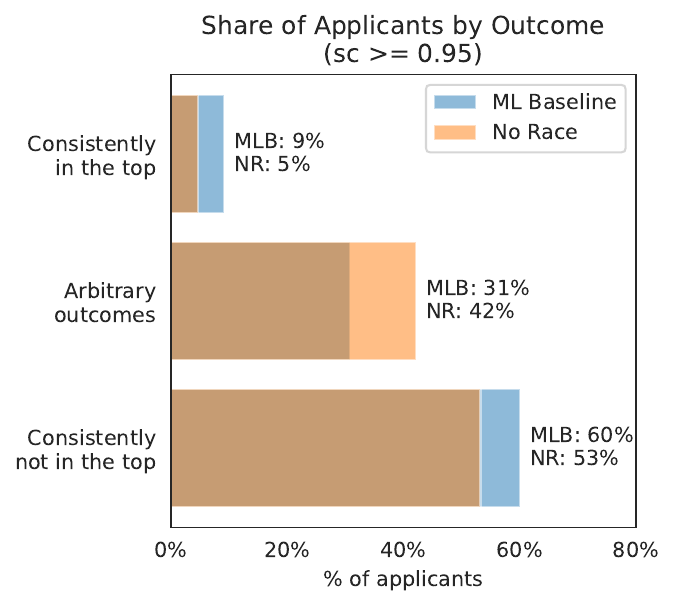}}
\Description{} 
\caption{Graph (a) shows the CDF of self-consistency for all applicants within 1,000 bootstraps of the ML baseline model for the applicant pool (blue line), within 1,000 bootstraps of the `no race' model (orange line), and across 500 bootstraps of each models (green line). Graph (b) shows CDFs only for those applicants who are usually top-ranked (ranked in the top by >50\% of bootstrapped models). Graph (c) shows the level of arbitrariness in the ML baseline model compared to the `No race' model, if we define an applicant’s outcomes to be consistent if and only if their \textsc{sc} $\geq$ 0.95: 9\% of applicants are consistently ranked in the top under the ML baseline model, and 5\% are consistently ranked in the top under the `No race' policy. 
\label{fig-across-within}}
\end{figure*}

The ML baseline model is prohibited under the SFFA policy change because it explicitly considers applicants' race and ethnicity as a feature in prioritizing them for review. A reasonable alternative to the ML baseline model is the `No race' model, which is identical but for the fact that it does not consider race as a feature. Fig.~\ref{fig-across-within}(a) shows self-consistency within the ML baseline model (blue curve) and within the 'No Race' policy (orange curve). Compared to the ML baseline model, the `No race' model has lower self-consistency -- meaning that applicants may be even more susceptible to experiencing arbitrary outcomes as a result of inherent randomness after the SFFA policy change takes effect. Inherent randomness in the `No race' modeling process creates arbitrariness that is 39\% larger than the arbitrariness created by inherent randomness in the ML baseline modeling process ($\textsc{ar} = 1.39$; Wilcoxon signed-rank test: $p < 0.001$).\looseness=-1

To provide a concrete example, we again compare consistency in outcomes at \textsc{sc} $\geq$ 0.95 in Fig.~\ref{fig-across-within}(c). Under the `No race' model, 58\% of applicants have consistent outcomes (compared to 69\% under the ML baseline model), and 5\% of applicants are consistently ranked in the top pool (compared to 9\% under the ML baseline model). As Fig.~\ref{fig-across-within}(a) shows, this reduced consistency within the `No race' model holds across all self-consistency thresholds. Again recalling that the top-ranked pool consists of 20\% of the full applicant pool, this means that under the `No race' model, three-quarters of the top pool will consist of applicants who have been added to that pool somewhat arbitrarily. Overall, these results imply that \textbf{inherent randomness, introduced through choices such as how to split training and test data, will play an even larger role in determining application review order following the SFFA policy change.}\looseness=-1

\paragraph{Arbitrariness across policies is larger than arbitrariness within a single policy.} 

Fig.~\ref{fig-across-within}(a) further shows self-consistency across both the ML baseline model and the 'No Race' policy (green curve). We observe that while the 'No Race' policy exhibits more arbitrariness than the ML baseline, the arbitrariness across these two policies is even larger. Intuitively, this means that changing a policy (in this case, to comply with the SFFA ruling), changes the outcomes that an applicant has, \textit{even though their overall merit as an applicant does not change}. To quantify this more precisely, the overall arbitrariness ratio \textsc{ar} of the across-policy outcomes to outcomes within the ML baseline model is 1.66. This means that the policy change creates a level of arbitrariness that is 66\% higher than inherent randomness present in the ML baseline modeling process alone (Wilcoxon signed-rank test: $p < 0.001$).  The \textsc{ar} of the across-policy outcomes to outcomes within the `No race' model is 1.19 -- the policy change increases arbitrariness($p < 0.001$), beyond the `No race' model. The observed pattern is even more pronounced for usually-top-ranked applicants, as shown in Fig.~\ref{fig-across-within}(b). For these applicants, the arbitrariness ratios of across-policy outcomes to outcomes within the ML baseline and `No race' models are 1.86 and 1.26, respectively (both $p < 0.001$). This means that the increase in arbitrariness created by policy change is \textit{even higher} among applicants who are usually-top-ranked.\looseness=-1

\paragraph{Within-policy arbitrariness increases for specific groups of applicants under the SFFA policy change}

\begin{figure*}[h]
\centering
\includegraphics[height=5cm]{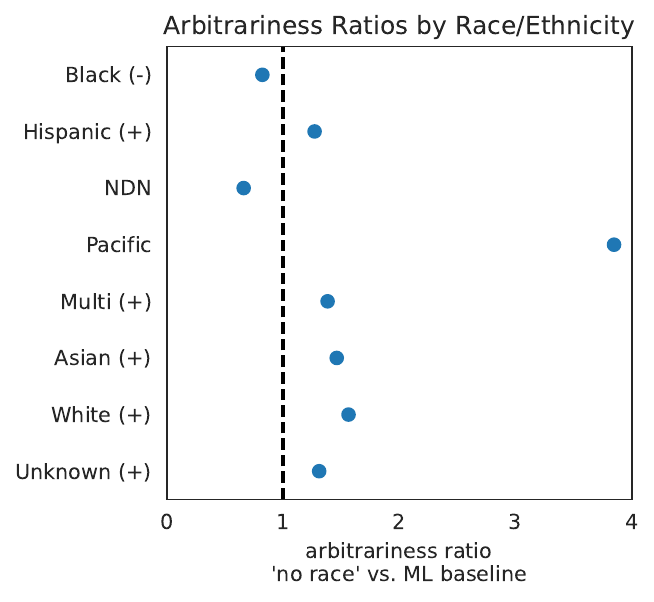}
\Description{} 
\caption{The arbitrariness ratios between the `No race' and the ML baseline policies for applicants of different races and ethnicities. A `+' denotes that arbitrariness for the group statistically significantly \textit{increases}, while a `-' denotes that arbitrariness for the group statistically significantly \textit{decreases}, per a Wilcoxon signed-rank test ($p < 0.001$ in all cases).
\label{fig-arbitrary-groups}}
\end{figure*}

Finally, we examine if the within-policy pattern of arbitrariness holds across racial and ethnic groups: as shown in Fig.~\ref{fig-arbitrary-groups}, arbitrariness in outcomes statistically significantly increases for Hispanic, multiracial, Asian, and White applicants, as well as applicants who did not report their race. Arbitrariness slightly decreases for Black applicants;\footnote{As implied by Fig.~\ref{fig-group-diversity-race}(a), this is due to the fact that Black applicants are more consistently \textit{not} ranked in the top under the `No race' model.} due to the under-representation of American Indian/Alaska Native and Native Hawaiian/Other Pacific Islander applicants in the pool, we are not able to observe statistically significant changes in arbitrariness for either group.\looseness=-1

\section{Discussion}\label{s-discussion}
In this study, we investigated how changes in admission policies brought about by the end of affirmative action are likely to impact applicant ranking algorithms, which are used to prioritize applications for review and which we argue can have a downstream impact on admissions decisions. We explored how hypothetical changes in admissions policies impact not only a college's overall class but also the outcomes of individual applicants.  To do this, we predicted and contextualized the likely impact of race-unaware admissions on the ability of colleges to admit a first-year class that displays both academic merit and diversity, building on impactful prior work conducting demographic fairness assessments in college admissions \cite{chetty_diversifying_2023, grossman_disparate_2023, marian_algorithmic_2023, zimdars_fairness_2010, robertson_modeling_2021}. We also provided a template for going beyond group fairness assessments to understand the relative impact of policy changes on \textit{individuals} by incorporating recent scholarship on model multiplicty and arbitrariness \cite{cooper2024arbitrariness, black_model_2022}.

We present four key findings. First, consistent with prior work \cite{lee2023evaluating}, we find that \textbf{race-unaware policies decrease the proportion of URM applicants represented in the top-ranked pool by 62\%}. Crucially, this change occurs \textbf{without a corresponding increase in academic merit of that top-ranked pool} (\S\ref{s-4.1}). Second, \textbf{omitting race from consideration reduces racial and socioeconomic diversity more than omitting other potentially informative variables, like applicants' intended major} (\S\ref{s-4.1}). Third, even in the absence of policy change, \textbf{inherent randomness in the modeling process will lead to somewhat arbitrary outcomes, especially for top-ranked applicants}: we find that across repeated bootstraps of the ML baseline model, just 9\% of applicants are consistently ranked in the top 20\%  (\S\ref{s-4.2}). Fourth, to contextualize the impact of policy change on individual applicant outcomes, we find that \textbf{arbitrariness resulting from both the policy change and inherent randomness in the modeling process is 66\% higher than arbitrariness resulting from inherent randomness alone} (\S\ref{s-4.2}). 
Lastly, \textbf{under a race-unaware applicant ranking algorithm, arbitrariness resulting from inherent randomness increases relative to the baseline for most applicants} (\S\ref{s-4.2}).\looseness=-1

In summary, our results imply that despite the impact of the SFFA policy change on college admissions processes,  complaints long attributed to affirmative action will persist at highly selective institutions; for example, many students with high test scores may not be ranked highly or ultimately admitted. We propose that this is because those complaints stem not from the specifics of any policy, including affirmative action, but from the fundamental issues of (1) limited space at selective American colleges and (2) inherent randomness in the admissions process. Because these constraints are intrinsic to the admissions system, we argue that ending affirmative action will not resolve these issues. 

\subsection{Ethical Considerations}
\paragraph{Adverse Impact.} Our aim with this work is to measure and contextualize challenges to admitting a diverse first-year class with high academic merit resulting from the SFFA policy change. However, we acknowledge that, by highlighting the modeling changes that most reduce the share of top-ranked URM, first-generation, and low-SES candidates, our work has the potential to be misused by those who wish to \textit{reduce} diversity in college admissions. However, because the changes we highlight are relatively simple (i.e. removal of sensitive variables), we argue that they are not likely to reveal previously unknown methods of discrimination, and therefore we believe the value in understanding the impact of the SFFA policy change outweighs the risk. 

\paragraph{Privacy and Human Subjects Research.} As part of this study, we had access to personal and sensitive information from college applicants. We took care to protect these data throughout the process of this study. All data was stored securely on dedicated servers that could only be accessed by approved individuals. All personally identifying information, including applicant names and contact information, were removed before we accessed and analyzed the data. This study was determined to be exempt by our institution's IRB. 

\subsection{Limitations}\label{s-limitations}
We acknowledge several limitations of our work. Chief among them is the narrow scope of our analysis. We focus on applicant ranking algorithms as one component of a larger admissions process and make an assumption that the order in which applicants are reviewed can impact admissions outcomes. While we believe that this is a reasonable assumption based on how researchers have previously described implementing admissions ranking algorithms in practice \cite{waters_grade_2014}, we are not able to precisely quantify this assumed impact. 

Our work is also narrow in another, larger sense: in examining the impact of the SFFA policy change on ranking algorithms only, we tacitly accept much of the status quo of college admissions. For example, we choose to follow prior work \cite{lee2023evaluating} and train our applicant ranking algorithms on \textit{past decisions}, effectively co-signing those as correctly identifying students with high academic merit (even though rejected students also may have had high merit). In order to mitigate this, we conducted a robustness assessment of an alternative target variable specification (Appendix \ref{a-3}), but we could have taken a more value-sensitive approach to applicant ranking algorithms instead (for example, by adopting a rule-based system that ranks applicants of lower socioeconomic status more highly even though the actual admissions process under-selects these students, as suggested by Chetty et al.~\cite{chetty_diversifying_2023}).\looseness=-1

We also chose to accept the Common App's definition of `under-represented minority candidate' and to assess diversity primarily according to that variable. Although our analysis showed that URM status is an important element to consider in the admissions process, essential questions about the significance of race and ethnicity relative to other uncontrollable applicant features (e.g. legacy, first-gen) remain. This focus is not merely specific to the institution in question but reflects broader societal and educational dynamics. Race often intersects with numerous other factors such as socioeconomic status, influencing higher educational opportunities and outcomes. Our study underscores the need to consider these intersections critically, recognizing race as a pivotal element in the complex matrix of college admissions.\looseness=-1

More broadly, with this work, we focus on \textit{computational solutions} to the \textit{sociotechnical problem} of bias and inequity in college admissions and thus forgo an examination of more transformational changes that the SFFA policy change could inspire \cite{abebe_roles_2020, green_escaping_2022}. However, we emphasize that our work is \textit{not prescriptive}; instead, we have sought to measure and contextualize likely changes to the applicant review process resulting from the SFFA policy change. We hope that our findings---that the SFFA policy change is likely to decrease the share of URM candidates who are given priority reviewing without meaningfully increasing the predictive power of ranking algorithms to identify applicants with high academic merit---will inspire future work to substantively improve racial, socioeconomic, and other forms of diversity in higher education.\looseness=-1

\subsection{Opportunities for Future Work}
By the time this work is published, data on the 2023-2024 college admissions cycle will be available, and researchers will be able to explore the extent to which our predictions on the impact of the SFFA policy change have come to pass, both in terms of how applicant ranking algorithms are modified and how the actually admitted class changes (and does not change). We believe that it will be important to conduct an empirical validation of our results, including an analysis into whether and why deviations from our predictions may occur (perhaps due to behavior changes from colleges and/or applicants in response to the SFFA policy change \cite{laufer_strategic_2023, liu2022strategic, hardt2016strategic, pmlr-v119-perdomo20a}). We also believe that it will be important to replicate our analysis across institutions to determine whether our findings hold across different applicant pools and admissions processes. In addition, our work suggests several avenues for future research. We show that, if previously built applicant ranking algorithms are minimally modified to be in compliance with the SFFA policy change, the share of top-ranked applicants who are URM is likely to fall by 62\%. It will therefore be important to identify alternative ranking approaches that can mitigate this impact, or that can increase other forms of diversity, like the share of top-ranked applicants that are first-generation or low-income (FGLI) students. Prior work related to equity and access in algorithms, including \citet{thomas_algorithmic_2021}'s framework for positive action, \citet{arif_khan_towards_2022}'s decision procedures for substantive equality of opportunity, and \citet{borgs_algorithmic_2019}'s approach to algorithmic greenlining, is likely to be instructive here; as will prior work on measuring and improving fairness without access to demographic data \cite{ashurst_fairness_2023}. We also show that arbitrariness can have a major impact on how applicants are ranked by ML models. In addition to considering arbitrariness in individual outcomes as a component of fairness assessments going forward, future work could explore how to reduce this arbitrariness, perhaps through bagging as suggested by \citet{cooper2024arbitrariness} or through other variance reduction techniques.\looseness=-1
\section{Conclusion}\label{s-conclusion}
In this work, we aim to quantify how changes in the admissions process for selective American colleges -- driven by a growing number of applications, test-optional policies, and the recent ban on race-conscious admissions -- will impact the order in which applicants are prioritized for review by admissions offices, and, by extension, who is admitted. We measure how the SFFA policy change will impact the overall applicant pool, finding that it is likely to reduce the share of top-ranked applicants who are URM without meaningfully increasing the academic merit of top-ranked applicants. We highlight that the omission of race has a more pronounced effect than the exclusion of other potentially informative but controllable features (e.g. major preference). Additionally, we explore how policy changes affect individual applicants, finding that inherent randomness in the modeling process will lead to somewhat arbitrary outcomes, especially for top-ranked applicants, and that arbitrariness is likely to increase as a result of the SFFA policy change.

\begin{acks}
This research was supported by a seed grant from the Cornell Center for Social Sciences. All content represents the opinion of the authors, which is not necessarily shared or endorsed by their respective employers and/or sponsors.
\end{acks}

\bibliographystyle{ACM-Reference-Format}
\bibliography{bibliography}


\begin{thebibliography}{73}


\ifx \showCODEN    \undefined \def \showCODEN     #1{\unskip}     \fi
\ifx \showDOI      \undefined \def \showDOI       #1{#1}\fi
\ifx \showISBNx    \undefined \def \showISBNx     #1{\unskip}     \fi
\ifx \showISBNxiii \undefined \def \showISBNxiii  #1{\unskip}     \fi
\ifx \showISSN     \undefined \def \showISSN      #1{\unskip}     \fi
\ifx \showLCCN     \undefined \def \showLCCN      #1{\unskip}     \fi
\ifx \shownote     \undefined \def \shownote      #1{#1}          \fi
\ifx \showarticletitle \undefined \def \showarticletitle #1{#1}   \fi
\ifx \showURL      \undefined \def \showURL       {\relax}        \fi
\providecommand\bibfield[2]{#2}
\providecommand\bibinfo[2]{#2}
\providecommand\natexlab[1]{#1}
\providecommand\showeprint[2][]{arXiv:#2}

\bibitem[sc1(1978)]%
        {sc1978bakke}
 \bibinfo{year}{1978}\natexlab{}.
\newblock \bibinfo{title}{University of California Regents v. Bakke}.
\newblock
\newblock
\newblock
\shownote{438 U.S. 265}.


\bibitem[sc2(2003a)]%
        {sc2003gratz}
 \bibinfo{year}{2003}\natexlab{a}.
\newblock \bibinfo{title}{Gratz v. Bollinger}.
\newblock
\newblock
\newblock
\shownote{539 U. S. 244}.


\bibitem[sc2(2003b)]%
        {sc2003grutter}
 \bibinfo{year}{2003}\natexlab{b}.
\newblock \bibinfo{title}{Grutter v. Bollinger}.
\newblock
\newblock
\newblock
\shownote{539 U. S. 306}.


\bibitem[sc2(2023)]%
        {sc2023students}
 \bibinfo{year}{2023}\natexlab{}.
\newblock \bibinfo{title}{Students for Fair Admissions, Inc. v. President and Fellows of Harvard College}.
\newblock
\newblock
\newblock
\shownote{600 U. S. \_\_\_}.


\bibitem[dar(2024)]%
        {dartmouthSAT}
 \bibinfo{year}{2024}\natexlab{}.
\newblock \bibinfo{title}{Report From Working Group on the Role of Standardized Test Scores in Undergraduate Admissions}.
\newblock
\newblock
\urldef\tempurl%
\url{https://home.dartmouth.edu/sites/home/files/2024-02/sat-undergrad-admissions.pdf}
\showURL{%
\tempurl}
\newblock
\shownote{Dartmouth College}.


\bibitem[Abebe et~al\mbox{.}(2020)]%
        {abebe_roles_2020}
\bibfield{author}{\bibinfo{person}{Rediet Abebe}, \bibinfo{person}{Solon Barocas}, \bibinfo{person}{Jon Kleinberg}, \bibinfo{person}{Karen Levy}, \bibinfo{person}{Manish Raghavan}, {and} \bibinfo{person}{David~G. Robinson}.} \bibinfo{year}{2020}\natexlab{}.
\newblock \showarticletitle{Roles for computing in social change}. In \bibinfo{booktitle}{\emph{Proceedings of the 2020 {Conference} on {Fairness}, {Accountability}, and {Transparency}}} \emph{(\bibinfo{series}{{FAT}*})}. \bibinfo{publisher}{Association for Computing Machinery}, \bibinfo{address}{New York, NY, USA}, \bibinfo{pages}{252--260}.
\newblock
\showISBNx{978-1-4503-6936-7}
\urldef\tempurl%
\url{https://doi.org/10.1145/3351095.3372871}
\showDOI{\tempurl}


\bibitem[Admissions({[n.\,d.]})]%
        {ColumbiaAdmissions}
\bibfield{author}{\bibinfo{person}{Columbia~Undergraduate Admissions}.} \bibinfo{year}{[n.\,d.]}\natexlab{}.
\newblock \bibinfo{title}{Testing Policy}.
\newblock
\newblock
\urldef\tempurl%
\url{https://undergrad.admissions.columbia.edu/apply/process/testing}
\showURL{%
\tempurl}


\bibitem[Admissions(2024)]%
        {MITadmissions}
\bibfield{author}{\bibinfo{person}{MIT Admissions}.} \bibinfo{year}{2024}\natexlab{}.
\newblock \bibinfo{title}{MIT Admissions}.
\newblock
\newblock
\urldef\tempurl%
\url{https://mitadmissions.org/}
\showURL{%
\tempurl}


\bibitem[Admissions(2023)]%
        {StanfordAdmissions}
\bibfield{author}{\bibinfo{person}{Stanford~Undergraduate Admissions}.} \bibinfo{year}{2023}\natexlab{}.
\newblock \bibinfo{title}{Admission Overview}.
\newblock
\newblock
\urldef\tempurl%
\url{https://admission.stanford.edu/apply/overview/index.html}
\showURL{%
\tempurl}


\bibitem[Alvero et~al\mbox{.}(2020)]%
        {alvero_ai_2020}
\bibfield{author}{\bibinfo{person}{A.J. Alvero}, \bibinfo{person}{Noah Arthurs}, \bibinfo{person}{anthony~lising antonio}, \bibinfo{person}{Benjamin~W. Domingue}, \bibinfo{person}{Ben Gebre-Medhin}, \bibinfo{person}{Sonia Giebel}, {and} \bibinfo{person}{Mitchell~L. Stevens}.} \bibinfo{year}{2020}\natexlab{}.
\newblock \showarticletitle{{AI} and {Holistic} {Review}: {Informing} {Human} {Reading} in {College} {Admissions}}. In \bibinfo{booktitle}{\emph{Proceedings of the {AAAI}/{ACM} {Conference} on {AI}, {Ethics}, and {Society}}} \emph{(\bibinfo{series}{{AIES}})}. \bibinfo{publisher}{Association for Computing Machinery}, \bibinfo{address}{New York, NY, USA}, \bibinfo{pages}{200--206}.
\newblock
\showISBNx{978-1-4503-7110-0}
\urldef\tempurl%
\url{https://doi.org/10.1145/3375627.3375871}
\showDOI{\tempurl}


\bibitem[Andrus and Villeneuve(2022)]%
        {andrus_demographic-reliant_2022}
\bibfield{author}{\bibinfo{person}{McKane Andrus} {and} \bibinfo{person}{Sarah Villeneuve}.} \bibinfo{year}{2022}\natexlab{}.
\newblock \showarticletitle{Demographic-{Reliant} {Algorithmic} {Fairness}: {Characterizing} the {Risks} of {Demographic} {Data} {Collection} in the {Pursuit} of {Fairness}}. In \bibinfo{booktitle}{\emph{2022 {ACM} {Conference} on {Fairness}, {Accountability}, and {Transparency}}} \emph{(\bibinfo{series}{{FAccT}})}. \bibinfo{publisher}{ACM}, \bibinfo{address}{Seoul Republic of Korea}, \bibinfo{pages}{1709--1721}.
\newblock
\showISBNx{978-1-4503-9352-2}
\urldef\tempurl%
\url{https://doi.org/10.1145/3531146.3533226}
\showDOI{\tempurl}


\bibitem[Arif~Khan et~al\mbox{.}(2022)]%
        {arif_khan_towards_2022}
\bibfield{author}{\bibinfo{person}{Falaah Arif~Khan}, \bibinfo{person}{Eleni Manis}, {and} \bibinfo{person}{Julia Stoyanovich}.} \bibinfo{year}{2022}\natexlab{}.
\newblock \showarticletitle{Towards {Substantive} {Conceptions} of {Algorithmic} {Fairness}: {Normative} {Guidance} from {Equal} {Opportunity} {Doctrines}}. In \bibinfo{booktitle}{\emph{Equity and {Access} in {Algorithms}, {Mechanisms}, and {Optimization}}} \emph{(\bibinfo{series}{{EAAMO}})}. \bibinfo{publisher}{ACM}, \bibinfo{address}{Arlington VA USA}, \bibinfo{pages}{1--10}.
\newblock
\showISBNx{978-1-4503-9477-2}
\urldef\tempurl%
\url{https://doi.org/10.1145/3551624.3555303}
\showDOI{\tempurl}


\bibitem[Arum and Stevens(2023)]%
        {nyt_selective}
\bibfield{author}{\bibinfo{person}{Richard Arum} {and} \bibinfo{person}{Mitchell~L. Stevens}.} \bibinfo{year}{2023}\natexlab{}.
\newblock \bibinfo{title}{For Most College Students, Affirmative Action Was Never Enough}.
\newblock
\newblock
\urldef\tempurl%
\url{https://www.nytimes.com/interactive/2023/07/03/opinion/for-most-college-students-affirmative-action-was-not-enough.html}
\showURL{%
\tempurl}
\newblock
\shownote{The New York Times}.


\bibitem[Ashurst and Weller(2023)]%
        {ashurst_fairness_2023}
\bibfield{author}{\bibinfo{person}{Carolyn Ashurst} {and} \bibinfo{person}{Adrian Weller}.} \bibinfo{year}{2023}\natexlab{}.
\newblock \showarticletitle{Fairness {Without} {Demographic} {Data}: {A} {Survey} of {Approaches}}. In \bibinfo{booktitle}{\emph{Proceedings of the 3rd {ACM} {Conference} on {Equity} and {Access} in {Algorithms}, {Mechanisms}, and {Optimization}}} \emph{(\bibinfo{series}{{EAAMO}})}. \bibinfo{publisher}{Association for Computing Machinery}, \bibinfo{address}{New York, NY, USA}, \bibinfo{pages}{1--12}.
\newblock
\showISBNx{9798400703812}
\urldef\tempurl%
\url{https://doi.org/10.1145/3617694.3623234}
\showDOI{\tempurl}


\bibitem[Asudeh et~al\mbox{.}(2019)]%
        {asudeh_2019_designing}
\bibfield{author}{\bibinfo{person}{Abolfazl Asudeh}, \bibinfo{person}{H.~V. Jagadish}, \bibinfo{person}{Julia Stoyanovich}, {and} \bibinfo{person}{Gautam Das}.} \bibinfo{year}{2019}\natexlab{}.
\newblock \showarticletitle{Designing Fair Ranking Schemes}. In \bibinfo{booktitle}{\emph{Proceedings of the 2019 International Conference on Management of Data}} (Amsterdam, Netherlands) \emph{(\bibinfo{series}{SIGMOD '19})}. \bibinfo{publisher}{Association for Computing Machinery}, \bibinfo{address}{New York, NY, USA}, \bibinfo{pages}{1259–1276}.
\newblock
\showISBNx{9781450356435}
\urldef\tempurl%
\url{https://doi.org/10.1145/3299869.3300079}
\showDOI{\tempurl}


\bibitem[Black et~al\mbox{.}(2022)]%
        {black_model_2022}
\bibfield{author}{\bibinfo{person}{Emily Black}, \bibinfo{person}{Manish Raghavan}, {and} \bibinfo{person}{Solon Barocas}.} \bibinfo{year}{2022}\natexlab{}.
\newblock \showarticletitle{Model {Multiplicity}: {Opportunities}, {Concerns}, and {Solutions}}. In \bibinfo{booktitle}{\emph{2022 {ACM} {Conference} on {Fairness}, {Accountability}, and {Transparency}}} \emph{(\bibinfo{series}{{FAccT}})}. \bibinfo{publisher}{ACM}, \bibinfo{address}{Seoul Republic of Korea}, \bibinfo{pages}{850--863}.
\newblock
\showISBNx{978-1-4503-9352-2}
\urldef\tempurl%
\url{https://doi.org/10.1145/3531146.3533149}
\showDOI{\tempurl}


\bibitem[Borgs et~al\mbox{.}(2019)]%
        {borgs_algorithmic_2019}
\bibfield{author}{\bibinfo{person}{Christian Borgs}, \bibinfo{person}{Jennifer Chayes}, \bibinfo{person}{Nika Haghtalab}, \bibinfo{person}{Adam~Tauman Kalai}, {and} \bibinfo{person}{Ellen Vitercik}.} \bibinfo{year}{2019}\natexlab{}.
\newblock \showarticletitle{Algorithmic {Greenlining}: {An} {Approach} to {Increase} {Diversity}}. In \bibinfo{booktitle}{\emph{Proceedings of the 2019 {AAAI}/{ACM} {Conference} on {AI}, {Ethics}, and {Society}}} \emph{(\bibinfo{series}{{AIES}})}. \bibinfo{publisher}{Association for Computing Machinery}, \bibinfo{address}{New York, NY, USA}, \bibinfo{pages}{69--76}.
\newblock
\showISBNx{978-1-4503-6324-2}
\urldef\tempurl%
\url{https://doi.org/10.1145/3306618.3314246}
\showDOI{\tempurl}


\bibitem[Bowen(1977)]%
        {princeton1977admissions}
\bibfield{author}{\bibinfo{person}{William~G. Bowen}.} \bibinfo{year}{1977}\natexlab{}.
\newblock \bibinfo{title}{Admissions and the Relevance of Race: Addressing the issues of principle, policy, and practice raised by the Bakke case}.
\newblock
\newblock
\urldef\tempurl%
\url{https://www.princeton.edu/~paw/web_exclusives/more/article_archives_bowen.html}
\showURL{%
\tempurl}
\newblock
\shownote{Princeton Alumni Weekly}.


\bibitem[Burke(2020)]%
        {IEDGRADE}
\bibfield{author}{\bibinfo{person}{Lilah Burke}.} \bibinfo{year}{2020}\natexlab{}.
\newblock \bibinfo{title}{The Death and Life of an Admissions Algorithm}.
\newblock
\newblock
\urldef\tempurl%
\url{https://www.insidehighered.com/admissions/article/2020/12/14/u-texas-will-stop-using-controversial-algorithm-evaluate-phd}
\showURL{%
\tempurl}
\newblock
\shownote{Inside Higher Ed}.


\bibitem[Celis et~al\mbox{.}(2018)]%
        {celis2018ranking}
\bibfield{author}{\bibinfo{person}{L.~Elisa Celis}, \bibinfo{person}{Damian Straszak}, {and} \bibinfo{person}{Nisheeth~K. Vishnoi}.} \bibinfo{year}{2018}\natexlab{}.
\newblock \bibinfo{title}{Ranking with Fairness Constraints}.
\newblock
\newblock
\showeprint[arxiv]{1704.06840}~[cs.DS]


\bibitem[Chetty et~al\mbox{.}(2023)]%
        {chetty_diversifying_2023}
\bibfield{author}{\bibinfo{person}{Raj Chetty}, \bibinfo{person}{David Deming}, {and} \bibinfo{person}{John Friedman}.} \bibinfo{year}{2023}\natexlab{}.
\newblock \bibinfo{booktitle}{\emph{Diversifying {Society}’s {Leaders}? {The} {Determinants} and {Causal} {Effects} of {Admission} to {Highly} {Selective} {Private} {Colleges}}}.
\newblock \bibinfo{type}{{T}echnical {R}eport} w31492. \bibinfo{institution}{National Bureau of Economic Research}, \bibinfo{address}{Cambridge, MA}. \bibinfo{pages}{w31492} pages.
\newblock
\urldef\tempurl%
\url{https://doi.org/10.3386/w31492}
\showDOI{\tempurl}


\bibitem[Coleman and Keith(2018)]%
        {collegeboard2018}
\bibfield{author}{\bibinfo{person}{Art Coleman} {and} \bibinfo{person}{Jamie~Lewis Keith}.} \bibinfo{year}{2018}\natexlab{}.
\newblock \bibinfo{title}{Understanding Holistic Review in Higher Education Admissions: Guiding Principles and Model Illustrations}.
\newblock
\newblock
\urldef\tempurl%
\url{https://highered.collegeboard.org/media/pdf/understanding-holistic-review-he-admissions.pdf}
\showURL{%
\tempurl}
\newblock
\shownote{College Board and EducationCounsel}.


\bibitem[Cooper et~al\mbox{.}(2024)]%
        {cooper2024arbitrariness}
\bibfield{author}{\bibinfo{person}{A.~Feder Cooper}, \bibinfo{person}{Katherine Lee}, \bibinfo{person}{Madiha~Zahrah Choksi}, \bibinfo{person}{Solon Barocas}, \bibinfo{person}{Christopher~De Sa}, \bibinfo{person}{James Grimmelmann}, \bibinfo{person}{Jon Kleinberg}, \bibinfo{person}{Siddhartha Sen}, {and} \bibinfo{person}{Baobao Zhang}.} \bibinfo{year}{2024}\natexlab{}.
\newblock \bibinfo{title}{Arbitrariness and Social Prediction: The Confounding Role of Variance in Fair Classification}.
\newblock
\newblock
\showeprint{2301.11562}


\bibitem[{Dutt} et~al\mbox{.}(2016)]%
        {dutt_gender_2016}
\bibfield{author}{\bibinfo{person}{Kuheli {Dutt}}, \bibinfo{person}{Danielle~L. {Pfaff}}, \bibinfo{person}{Ariel~F. {Bernstein}}, \bibinfo{person}{Joseph~S. {Dillard}}, {and} \bibinfo{person}{Caryn~J. {Block}}.} \bibinfo{year}{2016}\natexlab{}.
\newblock \showarticletitle{{Gender differences in recommendation letters for postdoctoral fellowships in geoscience}}.
\newblock \bibinfo{journal}{\emph{Nature Geoscience}} \bibinfo{volume}{9}, \bibinfo{number}{11} (\bibinfo{date}{Nov.} \bibinfo{year}{2016}), \bibinfo{pages}{805--808}.
\newblock
\urldef\tempurl%
\url{https://doi.org/10.1038/ngeo2819}
\showDOI{\tempurl}


\bibitem[Friedman et~al\mbox{.}(2024)]%
        {friedman_2024_standardized}
\bibfield{author}{\bibinfo{person}{John Friedman}, \bibinfo{person}{Bruce Sacerdote}, {and} \bibinfo{person}{Michele Tine}.} \bibinfo{year}{2024}\natexlab{}.
\newblock \bibinfo{title}{Standardized Test Scores and Academic Performance at Ivy-Plus Colleges}.
\newblock
\newblock
\urldef\tempurl%
\url{https://opportunityinsights.org/wp-content/uploads/2024/01/SAT_ACT_on_Grades.pdf}
\showURL{%
\tempurl}
\newblock
\shownote{Opportunity Insights}.


\bibitem[Garg et~al\mbox{.}(2021)]%
        {garg_standardized_2021}
\bibfield{author}{\bibinfo{person}{Nikhil Garg}, \bibinfo{person}{Hannah Li}, {and} \bibinfo{person}{Faidra Monachou}.} \bibinfo{year}{2021}\natexlab{}.
\newblock \showarticletitle{Standardized {Tests} and {Affirmative} {Action}: {The} {Role} of {Bias} and {Variance}}. In \bibinfo{booktitle}{\emph{Proceedings of the 2021 {ACM} {Conference} on {Fairness}, {Accountability}, and {Transparency}}} \emph{(\bibinfo{series}{{FAccT}})}. \bibinfo{publisher}{Association for Computing Machinery}, \bibinfo{address}{New York, NY, USA}, \bibinfo{pages}{261}.
\newblock
\showISBNx{978-1-4503-8309-7}
\urldef\tempurl%
\url{https://doi.org/10.1145/3442188.3445889}
\showDOI{\tempurl}


\bibitem[Green(2022)]%
        {green_escaping_2022}
\bibfield{author}{\bibinfo{person}{Ben Green}.} \bibinfo{year}{2022}\natexlab{}.
\newblock \showarticletitle{Escaping the {Impossibility} of {Fairness}: {From} {Formal} to {Substantive} {Algorithmic} {Fairness}}.
\newblock \bibinfo{journal}{\emph{Philosophy \& Technology}} \bibinfo{volume}{35}, \bibinfo{number}{4} (\bibinfo{date}{Dec.} \bibinfo{year}{2022}), \bibinfo{pages}{90}.
\newblock
\showISSN{2210-5433, 2210-5441}
\urldef\tempurl%
\url{https://doi.org/10.1007/s13347-022-00584-6}
\showDOI{\tempurl}


\bibitem[Grossman et~al\mbox{.}(2023)]%
        {grossman_disparate_2023}
\bibfield{author}{\bibinfo{person}{Joshua Grossman}, \bibinfo{person}{Sabina Tomkins}, \bibinfo{person}{Lindsay Page}, {and} \bibinfo{person}{Sharad Goel}.} \bibinfo{year}{2023}\natexlab{}.
\newblock \bibinfo{booktitle}{\emph{The {Disparate} {Impacts} of {College} {Admissions} {Policies} on {Asian} {American} {Applicants}}}.
\newblock \bibinfo{type}{{T}echnical {R}eport} w31527. \bibinfo{institution}{National Bureau of Economic Research}, \bibinfo{address}{Cambridge, MA}. \bibinfo{pages}{w31527} pages.
\newblock
\urldef\tempurl%
\url{https://doi.org/10.3386/w31527}
\showDOI{\tempurl}


\bibitem[Hardt et~al\mbox{.}(2016)]%
        {hardt2016strategic}
\bibfield{author}{\bibinfo{person}{Moritz Hardt}, \bibinfo{person}{Nimrod Megiddo}, \bibinfo{person}{Christos Papadimitriou}, {and} \bibinfo{person}{Mary Wootters}.} \bibinfo{year}{2016}\natexlab{}.
\newblock \showarticletitle{Strategic classification}. In \bibinfo{booktitle}{\emph{Proceedings of the 2016 ACM conference on innovations in theoretical computer science}}. \bibinfo{pages}{111--122}.
\newblock


\bibitem[Hartocollis and Harmon(2023)]%
        {nyt_affirmative}
\bibfield{author}{\bibinfo{person}{Anemona Hartocollis} {and} \bibinfo{person}{Amy Harmon}.} \bibinfo{year}{2023}\natexlab{}.
\newblock \bibinfo{title}{Affirmative Action Ruling Shakes Universities Over More Than Race}.
\newblock
\newblock
\urldef\tempurl%
\url{https://www.nytimes.com/2023/07/26/us/affirmative-action-college-admissions-harvard.html}
\showURL{%
\tempurl}
\newblock
\shownote{The New York Times}.


\bibitem[Huynh et~al\mbox{.}(2024)]%
        {huynh_2024_mitigating}
\bibfield{author}{\bibinfo{person}{Benjamin~Q. Huynh}, \bibinfo{person}{Elizabeth~T. Chin}, \bibinfo{person}{Allison Koenecke}, \bibinfo{person}{Derek Ouyang}, \bibinfo{person}{Daniel~E. Ho}, \bibinfo{person}{Mathew~V. Kiang}, {and} \bibinfo{person}{David~H. Rehkopf}.} \bibinfo{year}{2024}\natexlab{}.
\newblock \showarticletitle{Mitigating allocative tradeoffs and harms in an environmental justice data tool}.
\newblock \bibinfo{journal}{\emph{Nature Machine Intelligence}} \bibinfo{volume}{6}, \bibinfo{number}{2} (\bibinfo{date}{01 Feb} \bibinfo{year}{2024}), \bibinfo{pages}{187--194}.
\newblock
\showISSN{2522-5839}
\urldef\tempurl%
\url{https://doi.org/10.1038/s42256-024-00793-y}
\showDOI{\tempurl}


\bibitem[Initiative(2023)]%
        {CDS}
\bibfield{author}{\bibinfo{person}{Common Data~Set Initiative}.} \bibinfo{year}{2023}\natexlab{}.
\newblock \bibinfo{title}{Common Data Set 2023-2024}.
\newblock
\newblock
\urldef\tempurl%
\url{https://commondataset.org/wp-content/uploads/2024/01/CDS_2023-2024-with-auto-sum.pdf}
\showURL{%
\tempurl}


\bibitem[Jacobs and Wallach(2021)]%
        {jacobs_measurement_2021}
\bibfield{author}{\bibinfo{person}{Abigail~Z. Jacobs} {and} \bibinfo{person}{Hanna Wallach}.} \bibinfo{year}{2021}\natexlab{}.
\newblock \showarticletitle{Measurement and {Fairness}}. In \bibinfo{booktitle}{\emph{Proceedings of the 2021 {ACM} {Conference} on {Fairness}, {Accountability}, and {Transparency}}} \emph{(\bibinfo{series}{{FAccT}})}. \bibinfo{publisher}{Association for Computing Machinery}, \bibinfo{address}{New York, NY, USA}, \bibinfo{pages}{375--385}.
\newblock
\showISBNx{978-1-4503-8309-7}
\urldef\tempurl%
\url{https://doi.org/10.1145/3442188.3445901}
\showDOI{\tempurl}


\bibitem[Kim et~al\mbox{.}(2024)]%
        {common2024deadline}
\bibfield{author}{\bibinfo{person}{Brian~Heseung Kim}, \bibinfo{person}{Elyse Armstrong}, \bibinfo{person}{Laurel Eckhouse}, \bibinfo{person}{Mark Freeman}, \bibinfo{person}{Rodney Hughes}, \bibinfo{person}{Trent Kajikawa}, {and} \bibinfo{person}{Michelle Sinofsky}.} \bibinfo{year}{2024}\natexlab{}.
\newblock \bibinfo{title}{Deadline updates, 2023–2024: First-year application trends through Feb 1}.
\newblock
\newblock
\urldef\tempurl%
\url{https://www.commonapp.org/files/Common-App-Deadline-Updates-2024.02.14.pdf}
\showURL{%
\tempurl}


\bibitem[Laufer et~al\mbox{.}(2023)]%
        {laufer_strategic_2023}
\bibfield{author}{\bibinfo{person}{Benjamin Laufer}, \bibinfo{person}{Jon Kleinberg}, \bibinfo{person}{Karen Levy}, {and} \bibinfo{person}{Helen Nissenbaum}.} \bibinfo{year}{2023}\natexlab{}.
\newblock \showarticletitle{Strategic {Evaluation}}. In \bibinfo{booktitle}{\emph{Proceedings of the 3rd {ACM} {Conference} on {Equity} and {Access} in {Algorithms}, {Mechanisms}, and {Optimization}}} \emph{(\bibinfo{series}{{EAAMO}})}. \bibinfo{publisher}{Association for Computing Machinery}, \bibinfo{address}{New York, NY, USA}, \bibinfo{pages}{1--12}.
\newblock
\showISBNx{9798400703812}
\urldef\tempurl%
\url{https://doi.org/10.1145/3617694.3623237}
\showDOI{\tempurl}


\bibitem[Lee et~al\mbox{.}(2023a)]%
        {lee2023evaluating}
\bibfield{author}{\bibinfo{person}{Hansol Lee}, \bibinfo{person}{Ren\'{e}~F. Kizilcec}, {and} \bibinfo{person}{Thorsten Joachims}.} \bibinfo{year}{2023}\natexlab{a}.
\newblock \showarticletitle{Evaluating a Learned Admission-Prediction Model as a Replacement for Standardized Tests in College Admissions}. In \bibinfo{booktitle}{\emph{Proceedings of the Tenth ACM Conference on Learning @ Scale}} (Copenhagen, Denmark) \emph{(\bibinfo{series}{L@S '23})}. \bibinfo{publisher}{Association for Computing Machinery}, \bibinfo{address}{New York, NY, USA}, \bibinfo{pages}{195–203}.
\newblock
\showISBNx{9798400700255}
\urldef\tempurl%
\url{https://doi.org/10.1145/3573051.3593382}
\showDOI{\tempurl}


\bibitem[Lee et~al\mbox{.}(2023b)]%
        {lee_augmenting_2023}
\bibfield{author}{\bibinfo{person}{Jinsook Lee}, \bibinfo{person}{Bradon Thymes}, \bibinfo{person}{Joyce Zhou}, \bibinfo{person}{Thorsten Joachims}, {and} \bibinfo{person}{Rene~F. Kizilcec}.} \bibinfo{year}{2023}\natexlab{b}.
\newblock \showarticletitle{Augmenting {Holistic} {Review} in {University} {Admission} using {Natural} {Language} {Processing} for {Essays} and {Recommendation} {Letters}}. In \bibinfo{booktitle}{\emph{Equity, {Diversity}, \& {Inclusion} in {Educational} {Technology} {Research} \& {Development} {Workshop}}} \emph{(\bibinfo{series}{{AIED}})}. \bibinfo{publisher}{arXiv}, \bibinfo{address}{Tokyo, Japan}.
\newblock
\urldef\tempurl%
\url{http://arxiv.org/abs/2306.17575}
\showURL{%
\tempurl}


\bibitem[Leonhardt(2024)]%
        {nyt2024misguided}
\bibfield{author}{\bibinfo{person}{David Leonhardt}.} \bibinfo{year}{2024}\natexlab{}.
\newblock \bibinfo{title}{The Misguided War on the SAT}.
\newblock
\newblock
\urldef\tempurl%
\url{https://www.nytimes.com/2024/01/07/briefing/the-misguided-war-on-the-sat.html}
\showURL{%
\tempurl}
\newblock
\shownote{The New York Times}.


\bibitem[Liu et~al\mbox{.}(2022)]%
        {liu2022strategic}
\bibfield{author}{\bibinfo{person}{Lydia~T Liu}, \bibinfo{person}{Nikhil Garg}, {and} \bibinfo{person}{Christian Borgs}.} \bibinfo{year}{2022}\natexlab{}.
\newblock \showarticletitle{Strategic ranking}. In \bibinfo{booktitle}{\emph{International Conference on Artificial Intelligence and Statistics}}. PMLR, \bibinfo{pages}{2489--2518}.
\newblock


\bibitem[Liu and Garg(2021)]%
        {liu2021test}
\bibfield{author}{\bibinfo{person}{Zhi Liu} {and} \bibinfo{person}{Nikhil Garg}.} \bibinfo{year}{2021}\natexlab{}.
\newblock \showarticletitle{Test-optional policies: Overcoming strategic behavior and informational gaps}. In \bibinfo{booktitle}{\emph{Proceedings of the 1st ACM Conference on Equity and Access in Algorithms, Mechanisms, and Optimization}}. \bibinfo{pages}{1--13}.
\newblock


\bibitem[Marian(2023)]%
        {marian_algorithmic_2023}
\bibfield{author}{\bibinfo{person}{Amelie Marian}.} \bibinfo{year}{2023}\natexlab{}.
\newblock \showarticletitle{Algorithmic {Transparency} and {Accountability} through {Crowdsourcing}: {A} {Study} of the {NYC} {School} {Admission} {Lottery}}. In \bibinfo{booktitle}{\emph{Proceedings of the 2023 {ACM} {Conference} on {Fairness}, {Accountability}, and {Transparency}}} \emph{(\bibinfo{series}{{FAccT}})}. \bibinfo{publisher}{Association for Computing Machinery}, \bibinfo{address}{New York, NY, USA}, \bibinfo{pages}{434--443}.
\newblock
\showISBNx{9798400701924}
\urldef\tempurl%
\url{https://doi.org/10.1145/3593013.3594009}
\showDOI{\tempurl}


\bibitem[McConvey et~al\mbox{.}(2023)]%
        {mcconvey_human-centered_2023}
\bibfield{author}{\bibinfo{person}{Kelly McConvey}, \bibinfo{person}{Shion Guha}, {and} \bibinfo{person}{Anastasia Kuzminykh}.} \bibinfo{year}{2023}\natexlab{}.
\newblock \showarticletitle{A {Human}-{Centered} {Review} of {Algorithms} in {Decision}-{Making} in {Higher} {Education}}. In \bibinfo{booktitle}{\emph{Proceedings of the 2023 {CHI} {Conference} on {Human} {Factors} in {Computing} {Systems}}} \emph{(\bibinfo{series}{{CHI}})}. \bibinfo{publisher}{ACM}, \bibinfo{address}{Hamburg Germany}, \bibinfo{pages}{1--15}.
\newblock
\showISBNx{978-1-4503-9421-5}
\urldef\tempurl%
\url{https://doi.org/10.1145/3544548.3580658}
\showDOI{\tempurl}


\bibitem[Meyer(2023)]%
        {brookings_end}
\bibfield{author}{\bibinfo{person}{Katharine Meyer}.} \bibinfo{year}{2023}\natexlab{}.
\newblock \bibinfo{title}{The end of race-conscious admissions}.
\newblock
\newblock
\urldef\tempurl%
\url{https://www.brookings.edu/articles/the-end-of-race-conscious-admissions/}
\showURL{%
\tempurl}
\newblock
\shownote{The Brookings Institution}.


\bibitem[Michael N.~Bastedo and Kelly(2018)]%
        {bastedo2018what}
\bibfield{author}{\bibinfo{person}{Kristen M.~Glasener Michael N.~Bastedo, Nicholas A.~Bowman} {and} \bibinfo{person}{Jandi~L. Kelly}.} \bibinfo{year}{2018}\natexlab{}.
\newblock \showarticletitle{What are We Talking About When We Talk About Holistic Review? Selective College Admissions and its Effects on Low-SES Students}.
\newblock \bibinfo{journal}{\emph{The Journal of Higher Education}} \bibinfo{volume}{89}, \bibinfo{number}{5} (\bibinfo{year}{2018}), \bibinfo{pages}{782--805}.
\newblock
\urldef\tempurl%
\url{https://doi.org/10.1080/00221546.2018.1442633}
\showDOI{\tempurl}


\bibitem[of~California Standardized Testing Task~Force(2020)]%
        {UCreport}
\bibfield{author}{\bibinfo{person}{University of California Standardized Testing Task~Force}.} \bibinfo{year}{2020}\natexlab{}.
\newblock \bibinfo{title}{Report of the UC Academic Council Standardized Testing Task Force}.
\newblock
\newblock
\urldef\tempurl%
\url{https://senate.universityofcalifornia.edu/_files/ underreview/sttf-report.pdf}
\showURL{%
\tempurl}


\bibitem[of~Education National Center~for Education~Statistics({[n.\,d.]})]%
        {NCES2023applications}
\bibfield{author}{\bibinfo{person}{U.S.~Department of~Education National Center~for Education~Statistics}.} \bibinfo{year}{[n.\,d.]}\natexlab{}.
\newblock \bibinfo{title}{Integrated Postsecondary Education Data System (IPEDS), Admissions component final data (fall 2014 - 2021) and provisional data (fall 2022)}.
\newblock
\newblock
\urldef\tempurl%
\url{https://nces.ed.gov/ipeds/TrendGenerator/app/answer/10/101}
\showURL{%
\tempurl}


\bibitem[of~Education Office~for Civil~Rights(2018)]%
        {DOEmath}
\bibfield{author}{\bibinfo{person}{U.S.~Department of~Education Office~for Civil~Rights}.} \bibinfo{year}{2018}\natexlab{}.
\newblock \bibinfo{title}{DATA HIGHLIGHTS ON SCIENCE, TECHNOLOGY, ENGINEERING, AND MATHEMATICS COURSE TAKING IN OUR NATION’S PUBLIC SCHOOLS}.
\newblock
\newblock
\urldef\tempurl%
\url{https://www2.ed.gov/about/offices/list/ocr/docs/stem-course-taking.pdf}
\showURL{%
\tempurl}
\newblock
\shownote{2015–16 CIVIL RIGHTS DATA COLLECTION STEM COURSE TAKING}.


\bibitem[of~the State~of Indiana(1843)]%
        {indianalaw}
\bibfield{author}{\bibinfo{person}{General~Assembly of~the State~of Indiana}.} \bibinfo{year}{1843}\natexlab{}.
\newblock \bibinfo{booktitle}{\emph{The revised statutes of the state of Indiana, passed at the twenty-seventh session of the General assembly}}.
\newblock \bibinfo{publisher}{J. Dowling and R. Cole, state printers}, \bibinfo{address}{Indianapolis.} 1154 pages.
\newblock
\urldef\tempurl%
\url{https://babel.hathitrust.org/cgi/pt?id=nyp.33433009073879&seq=352}
\showURL{%
\tempurl}
\newblock
\shownote{By an act of the General assembly Samuel Bigger was 'authorized to prepare a compilation and revision of the general statute laws'. George H. Dunn was associated with the reviser}.


\bibitem[of~the State~of North~Carolina(1830)]%
        {ncantiliteracy}
\bibfield{author}{\bibinfo{person}{General~Assembly of~the State~of North~Carolina}.} \bibinfo{year}{1830}\natexlab{}.
\newblock \bibinfo{title}{Acts passed by the General Assembly of the State of North Carolina (1830-1831)}.
\newblock
\newblock
\urldef\tempurl%
\url{https://digital.ncdcr.gov/Documents/Detail/acts-passed-by-the-general-assembly-of-the-state-of-north-carolina-1830-1831/1955764?item=2080405}
\showURL{%
\tempurl}
\newblock
\shownote{Chapter VI: A Bill to Prevent All Persons from Teaching Slaves to Read or Write, the Use of Figures Excepted (page 15)}.


\bibitem[on~Harvard \& the Legacy~of Slavery et~al\mbox{.}(2022)]%
        {harvardlegacy}
\bibfield{author}{\bibinfo{person}{Presidential~Committee on~Harvard \& the Legacy~of Slavery}, \bibinfo{person}{Tomiko Brown-Nagin}, \bibinfo{person}{Sven Beckert}, \bibinfo{person}{Annette Gordon-Reed}, \bibinfo{person}{Stephen Gray}, \bibinfo{person}{Evelynn~M. Hammonds}, \bibinfo{person}{Nancy~F. Koehn}, \bibinfo{person}{Meira Levinson}, \bibinfo{person}{Tiya Miles}, \bibinfo{person}{Martha Minow}, \bibinfo{person}{Maya Sen}, \bibinfo{person}{Daniel~Albert Smith}, \bibinfo{person}{David~R. Williams}, {and} \bibinfo{person}{William~Julius Wilson}.} \bibinfo{year}{2022}\natexlab{}.
\newblock \bibinfo{title}{Harvard \& the Legacy of Slavery}.
\newblock
\newblock
\urldef\tempurl%
\url{https://legacyofslavery.harvard.edu/}
\showURL{%
\tempurl}


\bibitem[Passi and Barocas(2019)]%
        {passi_problem_2019}
\bibfield{author}{\bibinfo{person}{Samir Passi} {and} \bibinfo{person}{Solon Barocas}.} \bibinfo{year}{2019}\natexlab{}.
\newblock \showarticletitle{Problem {Formulation} and {Fairness}}. In \bibinfo{booktitle}{\emph{Proceedings of the {Conference} on {Fairness}, {Accountability}, and {Transparency}}} \emph{(\bibinfo{series}{{FAT}*})}. \bibinfo{publisher}{Association for Computing Machinery}, \bibinfo{address}{New York, NY, USA}, \bibinfo{pages}{39--48}.
\newblock
\showISBNx{978-1-4503-6125-5}
\urldef\tempurl%
\url{https://doi.org/10.1145/3287560.3287567}
\showDOI{\tempurl}


\bibitem[Patro et~al\mbox{.}(2022)]%
        {patro2022fair}
\bibfield{author}{\bibinfo{person}{Gourab~K Patro}, \bibinfo{person}{Lorenzo Porcaro}, \bibinfo{person}{Laura Mitchell}, \bibinfo{person}{Qiuyue Zhang}, \bibinfo{person}{Meike Zehlike}, {and} \bibinfo{person}{Nikhil Garg}.} \bibinfo{year}{2022}\natexlab{}.
\newblock \showarticletitle{Fair ranking: a critical review, challenges, and future directions}. In \bibinfo{booktitle}{\emph{Proceedings of the 2022 ACM conference on fairness, accountability, and transparency}}. \bibinfo{pages}{1929--1942}.
\newblock


\bibitem[Perdomo et~al\mbox{.}(2020)]%
        {pmlr-v119-perdomo20a}
\bibfield{author}{\bibinfo{person}{Juan Perdomo}, \bibinfo{person}{Tijana Zrnic}, \bibinfo{person}{Celestine Mendler-D{\"u}nner}, {and} \bibinfo{person}{Moritz Hardt}.} \bibinfo{year}{2020}\natexlab{}.
\newblock \showarticletitle{Performative Prediction}. In \bibinfo{booktitle}{\emph{Proceedings of the 37th International Conference on Machine Learning}} \emph{(\bibinfo{series}{Proceedings of Machine Learning Research}, Vol.~\bibinfo{volume}{119})}, \bibfield{editor}{\bibinfo{person}{Hal~Daumé III} {and} \bibinfo{person}{Aarti Singh}} (Eds.). \bibinfo{publisher}{PMLR}, \bibinfo{pages}{7599--7609}.
\newblock
\urldef\tempurl%
\url{https://proceedings.mlr.press/v119/perdomo20a.html}
\showURL{%
\tempurl}


\bibitem[Ragab et~al\mbox{.}(2012)]%
        {ragab2012hrspca}
\bibfield{author}{\bibinfo{person}{Abdul Hamid~M Ragab}, \bibinfo{person}{Abdul Fatah~S Mashat}, {and} \bibinfo{person}{Ahmed~M Khedra}.} \bibinfo{year}{2012}\natexlab{}.
\newblock \showarticletitle{HRSPCA: Hybrid recommender system for predicting college admission}. In \bibinfo{booktitle}{\emph{2012 12th International conference on intelligent systems design and applications (ISDA)}}. IEEE, \bibinfo{pages}{107--113}.
\newblock


\bibitem[Reardon(2011)]%
        {reardon_widening_2011}
\bibfield{author}{\bibinfo{person}{S.F. Reardon}.} \bibinfo{year}{2011}\natexlab{}.
\newblock \bibinfo{title}{The widening academic achievement gap between the rich and the poor: New evidence and possible explanations}.
\newblock
\newblock
\urldef\tempurl%
\url{https://cepa.stanford.edu/sites/default/files/reardon%20whither%20opportunity%20-%20chapter%205.pdf}
\showURL{%
\tempurl}
\newblock
\shownote{In R. Murnane \& G. Duncan (Eds.), Whither Opportunity? Rising Inequality and the Uncertain Life Chances of Low-Income Children. New York: Russell Sage Foundation Press}.


\bibitem[Reber et~al\mbox{.}(2023)]%
        {brookings2023admissions}
\bibfield{author}{\bibinfo{person}{Sarah Reber}, \bibinfo{person}{Gabriela Goodman}, {and} \bibinfo{person}{Rina Nagashima}.} \bibinfo{year}{2023}\natexlab{}.
\newblock \bibinfo{title}{Admissions at most colleges will be unaffected by Supreme Court ruling on affirmative action}.
\newblock
\newblock
\urldef\tempurl%
\url{https://www.brookings.edu/articles/admissions-at-most-colleges-will-be-unaffected-by-supreme-court-ruling-on-affirmative-action/}
\showURL{%
\tempurl}
\newblock
\shownote{The Brookings Institution}.


\bibitem[Robertson et~al\mbox{.}(2021)]%
        {robertson_modeling_2021}
\bibfield{author}{\bibinfo{person}{Samantha Robertson}, \bibinfo{person}{Tonya Nguyen}, {and} \bibinfo{person}{Niloufar Salehi}.} \bibinfo{year}{2021}\natexlab{}.
\newblock \showarticletitle{Modeling {Assumptions} {Clash} with the {Real} {World}: {Transparency}, {Equity}, and {Community} {Challenges} for {Student} {Assignment} {Algorithms}}. In \bibinfo{booktitle}{\emph{Proceedings of the 2021 {CHI} {Conference} on {Human} {Factors} in {Computing} {Systems}}} \emph{(\bibinfo{series}{{CHI}})}. \bibinfo{publisher}{ACM}, \bibinfo{address}{Yokohama Japan}, \bibinfo{pages}{1--14}.
\newblock
\showISBNx{978-1-4503-8096-6}
\urldef\tempurl%
\url{https://doi.org/10.1145/3411764.3445748}
\showDOI{\tempurl}


\bibitem[Shao et~al\mbox{.}(2022)]%
        {shao2022combinatorial}
\bibfield{author}{\bibinfo{person}{Lucy Shao}, \bibinfo{person}{Richard~A. Levine}, \bibinfo{person}{Stefan Hyman}, \bibinfo{person}{Jeanne Stronach}, {and} \bibinfo{person}{Juanjuan Fan}.} \bibinfo{year}{2022}\natexlab{}.
\newblock \showarticletitle{A Combinatorial Optimization Framework for Scoring Students in University Admissions}.
\newblock \bibinfo{journal}{\emph{Evaluation Review}} \bibinfo{volume}{46}, \bibinfo{number}{3} (\bibinfo{year}{2022}), \bibinfo{pages}{296--335}.
\newblock
\urldef\tempurl%
\url{https://doi.org/10.1177/0193841X221082887}
\showDOI{\tempurl}


\bibitem[Singh and Joachims(2018)]%
        {singh_2018_fairness}
\bibfield{author}{\bibinfo{person}{Ashudeep Singh} {and} \bibinfo{person}{Thorsten Joachims}.} \bibinfo{year}{2018}\natexlab{}.
\newblock \showarticletitle{Fairness of Exposure in Rankings}. In \bibinfo{booktitle}{\emph{Proceedings of the 24th ACM SIGKDD International Conference on Knowledge Discovery \& Data Mining}} (London, United Kingdom) \emph{(\bibinfo{series}{KDD '18})}. \bibinfo{publisher}{Association for Computing Machinery}, \bibinfo{address}{New York, NY, USA}, \bibinfo{pages}{2219–2228}.
\newblock
\showISBNx{9781450355520}
\urldef\tempurl%
\url{https://doi.org/10.1145/3219819.3220088}
\showDOI{\tempurl}


\bibitem[Singh and Joachims(2019)]%
        {singh_2019_policy}
\bibfield{author}{\bibinfo{person}{Ashudeep Singh} {and} \bibinfo{person}{Thorsten Joachims}.} \bibinfo{year}{2019}\natexlab{}.
\newblock \showarticletitle{Policy Learning for Fairness in Ranking}. In \bibinfo{booktitle}{\emph{Advances in Neural Information Processing Systems}}, \bibfield{editor}{\bibinfo{person}{H.~Wallach}, \bibinfo{person}{H.~Larochelle}, \bibinfo{person}{A.~Beygelzimer}, \bibinfo{person}{F.~d\textquotesingle Alch\'{e}-Buc}, \bibinfo{person}{E.~Fox}, {and} \bibinfo{person}{R.~Garnett}} (Eds.), Vol.~\bibinfo{volume}{32}. \bibinfo{publisher}{Curran Associates, Inc.}
\newblock
\urldef\tempurl%
\url{https://proceedings.neurips.cc/paper_files/paper/2019/file/9e82757e9a1c12cb710ad680db11f6f1-Paper.pdf}
\showURL{%
\tempurl}


\bibitem[Sridhar et~al\mbox{.}(2020)]%
        {sridhar2020university}
\bibfield{author}{\bibinfo{person}{Sashank Sridhar}, \bibinfo{person}{Siddartha Mootha}, {and} \bibinfo{person}{Santosh Kolagati}.} \bibinfo{year}{2020}\natexlab{}.
\newblock \showarticletitle{A university admission prediction system using stacked ensemble learning}. In \bibinfo{booktitle}{\emph{2020 Advanced Computing and Communication Technologies for High Performance Applications (ACCTHPA)}}. IEEE, \bibinfo{pages}{162--167}.
\newblock


\bibitem[Staudaher et~al\mbox{.}(2020)]%
        {staudaher2020prediction}
\bibfield{author}{\bibinfo{person}{Shawn Staudaher}, \bibinfo{person}{Jeonghyun Lee}, {and} \bibinfo{person}{Farahnaz Soleimani}.} \bibinfo{year}{2020}\natexlab{}.
\newblock \showarticletitle{Predicting Applicant Admission Status for Georgia Tech's Online Master's in Analytics Program}. In \bibinfo{booktitle}{\emph{Proceedings of the Seventh ACM Conference on Learning @ Scale}} (Virtual Event, USA) \emph{(\bibinfo{series}{L@S '20})}. \bibinfo{publisher}{Association for Computing Machinery}, \bibinfo{address}{New York, NY, USA}, \bibinfo{pages}{309–312}.
\newblock
\showISBNx{9781450379519}
\urldef\tempurl%
\url{https://doi.org/10.1145/3386527.3406735}
\showDOI{\tempurl}


\bibitem[Stevens(2007)]%
        {stevens_2007_creating}
\bibfield{author}{\bibinfo{person}{Mitchell~L. Stevens}.} \bibinfo{year}{2007}\natexlab{}.
\newblock \bibinfo{booktitle}{\emph{Creating a Class: College Admissions and the Education of Elites}}.
\newblock \bibinfo{publisher}{Harvard University Press}.
\newblock
\showISBNx{9780674026735}
\urldef\tempurl%
\url{http://www.jstor.org/stable/j.ctv1m46g11}
\showURL{%
\tempurl}


\bibitem[Stulberg and Chen(2014)]%
        {stulberg2014origins}
\bibfield{author}{\bibinfo{person}{Lisa~M. Stulberg} {and} \bibinfo{person}{Anthony~S. Chen}.} \bibinfo{year}{2014}\natexlab{}.
\newblock \showarticletitle{The Origins of Race-conscious Affirmative Action in Undergraduate Admissions: A Comparative Analysis of Institutional Change in Higher Education}.
\newblock \bibinfo{journal}{\emph{Sociology of Education}} \bibinfo{volume}{87}, \bibinfo{number}{1} (\bibinfo{year}{2014}), \bibinfo{pages}{36--52}.
\newblock
\urldef\tempurl%
\url{https://doi.org/10.1177/0038040713514063}
\showDOI{\tempurl}


\bibitem[Thomas et~al\mbox{.}(2021)]%
        {thomas_algorithmic_2021}
\bibfield{author}{\bibinfo{person}{Oliver Thomas}, \bibinfo{person}{Miri Zilka}, \bibinfo{person}{Adrian Weller}, {and} \bibinfo{person}{Novi Quadrianto}.} \bibinfo{year}{2021}\natexlab{}.
\newblock \showarticletitle{An {Algorithmic} {Framework} for {Positive} {Action}}. In \bibinfo{booktitle}{\emph{Equity and {Access} in {Algorithms}, {Mechanisms}, and {Optimization}}} \emph{(\bibinfo{series}{{EAAMO}})}. \bibinfo{publisher}{ACM}, \bibinfo{address}{-- NY USA}, \bibinfo{pages}{1--13}.
\newblock
\showISBNx{978-1-4503-8553-4}
\urldef\tempurl%
\url{https://doi.org/10.1145/3465416.3483303}
\showDOI{\tempurl}


\bibitem[University(2024)]%
        {CMUadmissions}
\bibfield{author}{\bibinfo{person}{Carnegie~Mellon University}.} \bibinfo{year}{2024}\natexlab{}.
\newblock \bibinfo{title}{Admissions Consideration}.
\newblock
\newblock
\urldef\tempurl%
\url{https://www.cmu.edu/admission/admission/admission-consideration}
\showURL{%
\tempurl}


\bibitem[Vaghela and Sharma(2015)]%
        {vaghela2015students}
\bibfield{author}{\bibinfo{person}{Dineshkumar~B Vaghela} {and} \bibinfo{person}{Priyanka Sharma}.} \bibinfo{year}{2015}\natexlab{}.
\newblock \showarticletitle{Students' Admission Prediction using GRBST with Distributed Data Mining}.
\newblock \bibinfo{journal}{\emph{Communications on Applied Electronics}} \bibinfo{volume}{2}, \bibinfo{number}{1} (\bibinfo{year}{2015}), \bibinfo{pages}{15--19}.
\newblock


\bibitem[Waters and Miikkulainen(2014)]%
        {waters_grade_2014}
\bibfield{author}{\bibinfo{person}{Austin Waters} {and} \bibinfo{person}{Risto Miikkulainen}.} \bibinfo{year}{2014}\natexlab{}.
\newblock \showarticletitle{{GRADE}: {Machine}‐{Learning} {Support} for {Graduate} {Admissions}}.
\newblock \bibinfo{journal}{\emph{AI Magazine}} \bibinfo{volume}{35}, \bibinfo{number}{1} (\bibinfo{date}{March} \bibinfo{year}{2014}), \bibinfo{pages}{64--75}.
\newblock
\showISSN{0738-4602, 2371-9621}
\urldef\tempurl%
\url{https://doi.org/10.1609/aimag.v35i1.2504}
\showDOI{\tempurl}


\bibitem[Watson-Daniels et~al\mbox{.}(2023)]%
        {watsondaniels_multitarget_2023}
\bibfield{author}{\bibinfo{person}{Jamelle Watson-Daniels}, \bibinfo{person}{Solon Barocas}, \bibinfo{person}{Jake~M. Hofman}, {and} \bibinfo{person}{Alexandra Chouldechova}.} \bibinfo{year}{2023}\natexlab{}.
\newblock \showarticletitle{Multi-{Target} {Multiplicity}: {Flexibility} and {Fairness} in {Target} {Specification} under {Resource} {Constraints}}. In \bibinfo{booktitle}{\emph{Proceedings of the 2023 {ACM} {Conference} on {Fairness}, {Accountability}, and {Transparency}}} \emph{(\bibinfo{series}{{FAccT}})}. \bibinfo{publisher}{Association for Computing Machinery}, \bibinfo{address}{New York, NY, USA}, \bibinfo{pages}{297--311}.
\newblock
\showISBNx{9798400701924}
\urldef\tempurl%
\url{https://doi.org/10.1145/3593013.3593998}
\showDOI{\tempurl}


\bibitem[Zehlike et~al\mbox{.}(2017)]%
        {zehlike_2017_fair}
\bibfield{author}{\bibinfo{person}{Meike Zehlike}, \bibinfo{person}{Francesco Bonchi}, \bibinfo{person}{Carlos Castillo}, \bibinfo{person}{Sara Hajian}, \bibinfo{person}{Mohamed Megahed}, {and} \bibinfo{person}{Ricardo Baeza-Yates}.} \bibinfo{year}{2017}\natexlab{}.
\newblock \showarticletitle{FA*IR: A Fair Top-k Ranking Algorithm}. In \bibinfo{booktitle}{\emph{Proceedings of the 2017 ACM on Conference on Information and Knowledge Management}} (Singapore, Singapore) \emph{(\bibinfo{series}{CIKM '17})}. \bibinfo{publisher}{Association for Computing Machinery}, \bibinfo{address}{New York, NY, USA}, \bibinfo{pages}{1569–1578}.
\newblock
\showISBNx{9781450349185}
\urldef\tempurl%
\url{https://doi.org/10.1145/3132847.3132938}
\showDOI{\tempurl}


\bibitem[Zehlike et~al\mbox{.}(2022a)]%
        {zehlike_2022_fairness}
\bibfield{author}{\bibinfo{person}{Meike Zehlike}, \bibinfo{person}{Ke Yang}, {and} \bibinfo{person}{Julia Stoyanovich}.} \bibinfo{year}{2022}\natexlab{a}.
\newblock \showarticletitle{Fairness in Ranking, Part I: Score-Based Ranking}.
\newblock \bibinfo{journal}{\emph{ACM Comput. Surv.}} \bibinfo{volume}{55}, \bibinfo{number}{6}, Article \bibinfo{articleno}{118} (\bibinfo{date}{dec} \bibinfo{year}{2022}), \bibinfo{numpages}{36}~pages.
\newblock
\showISSN{0360-0300}
\urldef\tempurl%
\url{https://doi.org/10.1145/3533379}
\showDOI{\tempurl}


\bibitem[Zehlike et~al\mbox{.}(2022b)]%
        {zehlike_2022_fairnessII}
\bibfield{author}{\bibinfo{person}{Meike Zehlike}, \bibinfo{person}{Ke Yang}, {and} \bibinfo{person}{Julia Stoyanovich}.} \bibinfo{year}{2022}\natexlab{b}.
\newblock \showarticletitle{Fairness in Ranking, Part II: Learning-to-Rank and Recommender Systems}.
\newblock \bibinfo{journal}{\emph{ACM Comput. Surv.}} \bibinfo{volume}{55}, \bibinfo{number}{6}, Article \bibinfo{articleno}{117} (\bibinfo{date}{dec} \bibinfo{year}{2022}), \bibinfo{numpages}{41}~pages.
\newblock
\showISSN{0360-0300}
\urldef\tempurl%
\url{https://doi.org/10.1145/3533380}
\showDOI{\tempurl}


\bibitem[Zimdars(2010)]%
        {zimdars_fairness_2010}
\bibfield{author}{\bibinfo{person}{Anna Zimdars}.} \bibinfo{year}{2010}\natexlab{}.
\newblock \showarticletitle{Fairness and undergraduate admission: a qualitative exploration of admissions choices at the {University} of {Oxford}}.
\newblock \bibinfo{journal}{\emph{Oxford Review of Education}} \bibinfo{volume}{36}, \bibinfo{number}{3} (\bibinfo{date}{June} \bibinfo{year}{2010}), \bibinfo{pages}{307--323}.
\newblock
\showISSN{0305-4985, 1465-3915}
\urldef\tempurl%
\url{https://doi.org/10.1080/03054981003732286}
\showDOI{\tempurl}


\end{thebibliography}

\newpage
\appendix

\section{Admissions at Selective American Colleges}\label{a-1}

\begin{table*}[h]
\centering
\caption{Details of the admissions process at Ivy-Plus institutions (which we define following Chetty et al.~\cite{chetty_diversifying_2023}). All data comes from the most recently released version of the Common Data Set for each institution.}
\label{cds-data}
\begin{threeparttable}
\begin{tabular}{lccccc} 
\toprule
\multicolumn{1}{c}{\textbf{Institution}} & \textbf{\# Applicants} & \textbf{\# Admitted} & \textbf{\# Waitlist} & \textbf{Application Due Date} & \textbf{Notification Date} \\ 
\midrule
Brown University\tnote{1} & 50,649 & 2,562 & - & January 5 & Late March \\
Columbia University\tnote{2} & 60,374 & 2,255 & - & January 1 & April 1 \\
Cornell University\tnote{3} & 71,164 & 5,168 & 7,729 & January 2 & Early April \\
Dartmouth College\tnote{4} & 28,336 & 1,808 & 2,098 & January 3 & Early April \\
Duke University\tnote{5} & 49,523 & 2,911 & - & January 3 & April 1 \\
Harvard University\tnote{6} & 61,221 & 1,984 & - & January 1 & April 1 \\
\begin{tabular}[c]{@{}l@{}}Massachusetts Institute \\of Technology\tnote{7}\end{tabular} & 33,767 & 1,337 & 763 & January 1 & March 20 \\
Princeton University\tnote{8} & 38,019 & 2,167 & 1,710 & January 1 & April 1 \\
Stanford University\tnote{9} & 56,378 & 2,075 & 553 & January 5 & April 1 \\
University of Chicago\tnote{10} & 37,974 & 2,460 & - & - & - \\
\begin{tabular}[c]{@{}l@{}}University of \\Pennsylvania\tnote{11}\end{tabular} & 54,588 & 3,549 & 3,351 & January 5 & April 1 \\
Yale University\tnote{12} & 50,060 & 2,289 & 1,000 & January 2 & April 1 \\
\bottomrule
\end{tabular}
\begin{tablenotes}\footnotesize
\item[1] \url{https://oir.brown.edu/sites/default/files/2020-04/CDS\_2022\_2023.pdf}
\item[2] \url{https://opir.columbia.edu/sites/default/files/content/Common\%20Data\%20Set/CDS\%20College\%20Engineering\%202022-2023.pdf}
\item[3] \url{https://irp.dpb.cornell.edu/wp-content/uploads/2023/09/CDS\_2022-2023\_Cornell-University-v7.pdf}
\item[4] \url{https://www.dartmouth.edu/oir/pdfs/cds\_2022-2023.pdf}
\item[5] \url{https://provost-files.cloud.duke.edu/sites/default/files/CDS\%202021-22\%20FINAL\_2.pdf}
\item[6] \url{https://bpb-us-e1.wpmucdn.com/sites.harvard.edu/dist/6/210/files/2023/06/harvard\_cds\_2022-2023.pdf}
\item[7] \url{https://ir.mit.edu/cds-2023}
\item[8] \url{https://registrar.princeton.edu/sites/g/files/toruqf136/files/documents/CDS\_2022-2023.pdf}
\item[9] \url{https://ucomm.stanford.edu/wp-content/uploads/sites/15/2023/03/CDS\_2022-2023\_v3.pdf}
\item[10] \url{https://bpb-us-w2.wpmucdn.com/voices.uchicago.edu/dist/8/2077/files/2022/10/UChicago\_CDS\_2021-22.pdf}
\item[11] \url{https://upenn.app.box.com/s/75jr7yip7279rcsfic0946o1pkr8okrt}
\item[12] \url{https://oir.yale.edu/sites/default/files/cds\_yale\_2022-2023\_vf\_10062023.pdf}
\end{tablenotes}
\end{threeparttable}
\end{table*}

\section{Robustness to Different `Top Pool' Cutoffs}\label{a-2}
\begin{figure*}[h]
\centering
\subfloat[]{\includegraphics[height=4.5cm]{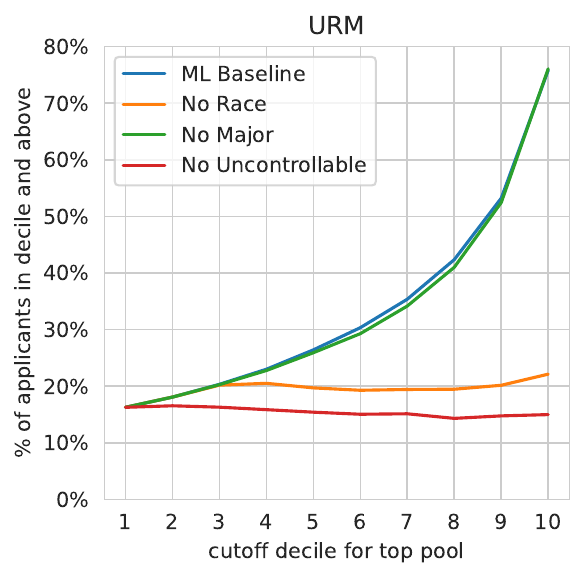}}
\hspace{0.05\textwidth}
\subfloat[]{\includegraphics[height=4.5cm]{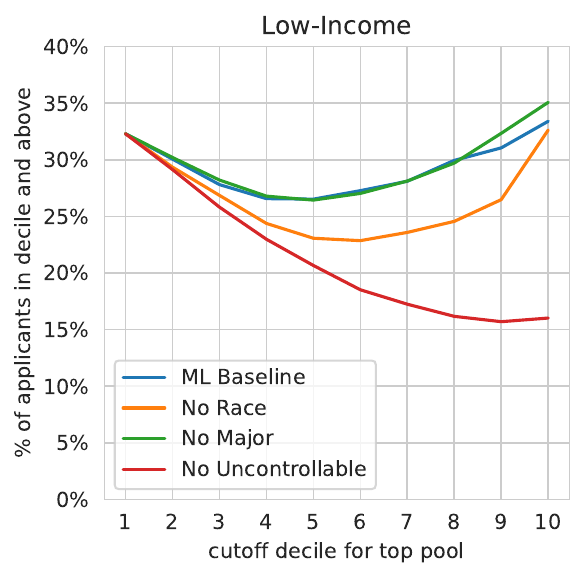}}
\hspace{0.05\textwidth}
\subfloat[]{\includegraphics[height=4.5cm]{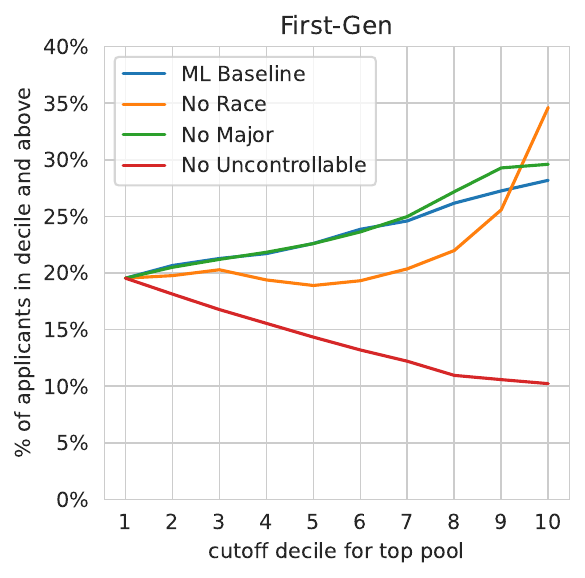}}
\hspace{0.05\textwidth}
\subfloat[]{\includegraphics[height=4.5cm]{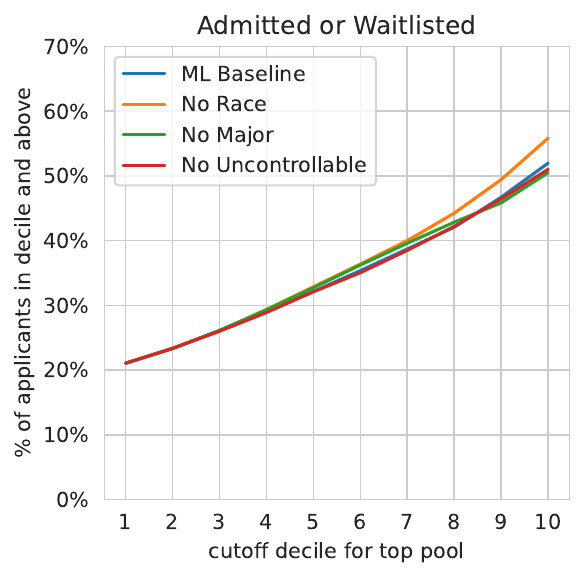}}
\hspace{0.05\textwidth}
\subfloat[]{\includegraphics[height=4.5cm]{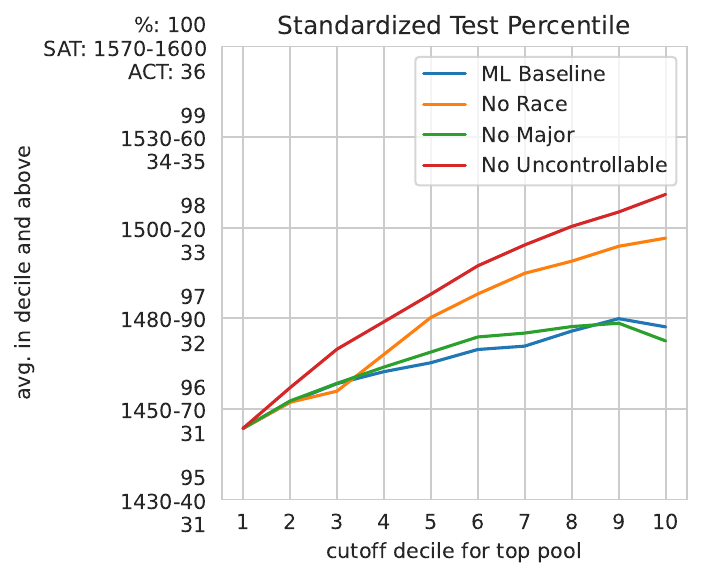}}
\Description{} 
\caption{Graphs (a), (b), (c), and (d) show the share of applicants in the top pool who belong to each of the specified groups as the `cutoff' (the minimum decile considered part of the top group) changes. Graph (e) shows the average standardized test percentile submitted by applicants who belong to the top group as the cutoff changes. Note that a cutoff value of 9 corresponds to the top group as defined in \S\ref{s-results}.
\label{fig-appendix-robustness}}
\end{figure*}

To ensure that our results are not brittle with respect to our specific choice of how to define the `top' pool of applicants, we conducted a robustness assessment, examining to what extent our findings about diversity and academic merit of the top-ranked pool change as the top pool itself changes. To do this, we vary the `cutoff' for the top pool: the minimum decile considered part of the top. As Fig.~\ref{fig-appendix-robustness} shows, the relative ordering of attributes across models remains constant at almost all cutoff choices. For example, the simulated policy changes that result in a decreased URM share within the top pool of applicants relative to the ML baseline model (the `no race' and `no uncontrollable features' models) do so whether the cutoff is set at Decile 10, Decile 9, Decile 8, Decile 7, and so on. The only exception to this is that the share of first-gen applicants included in the top pool by the `no race' model \textit{increases} relative to the ML baseline if the cutoff is set at Decile 10, but \textit{decreases} relative to the ML baseline if the cutoff is set at any other decile. The results we discuss in \S\ref{s-results} use Decile 9 as the cutoff, which is consistent with most other cutoffs. Additionally, the magnitude of differences between models decreases as the cutoff decile decreases. This makes intuitive sense, as it indicates that a higher proportion of the overall applicant pool is included in the `top' pool. When the cutoff is Decile 1, all applicants are included in the top pool, and the demographic and academic features of the top applicants converge to the averages of the test set. 

\section{Robustness to Target Variable Selection}\label{a-3}

\begin{figure*}[!h]
\centering
\subfloat[]{\includegraphics[height=4.5cm]{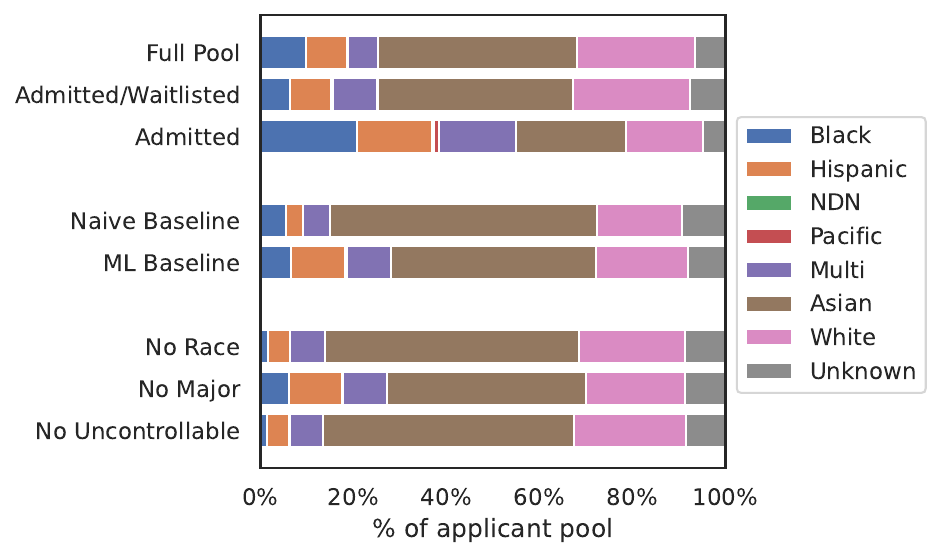}}
\hspace{0.05\textwidth}
\subfloat[]{\includegraphics[height=4.5cm]{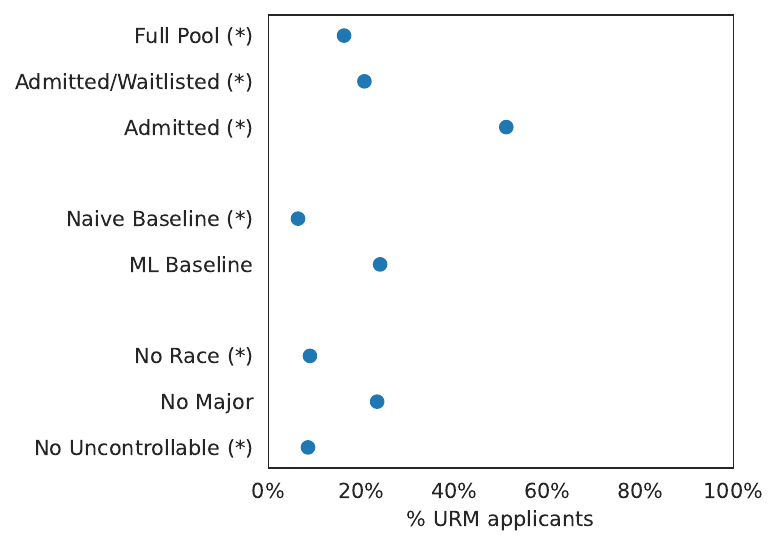}}
\hspace{0.05\textwidth}
\subfloat[]{\includegraphics[height=4.5cm]{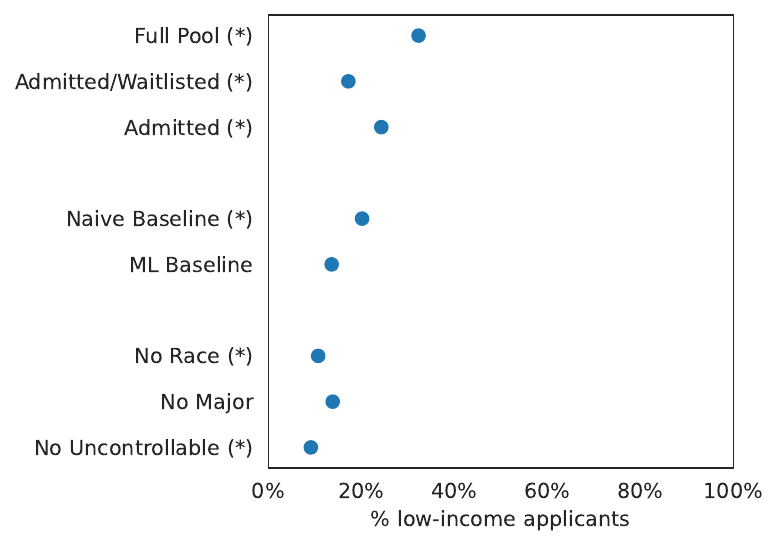}}
\hspace{0.05\textwidth}
\subfloat[]{\includegraphics[height=4.5cm]{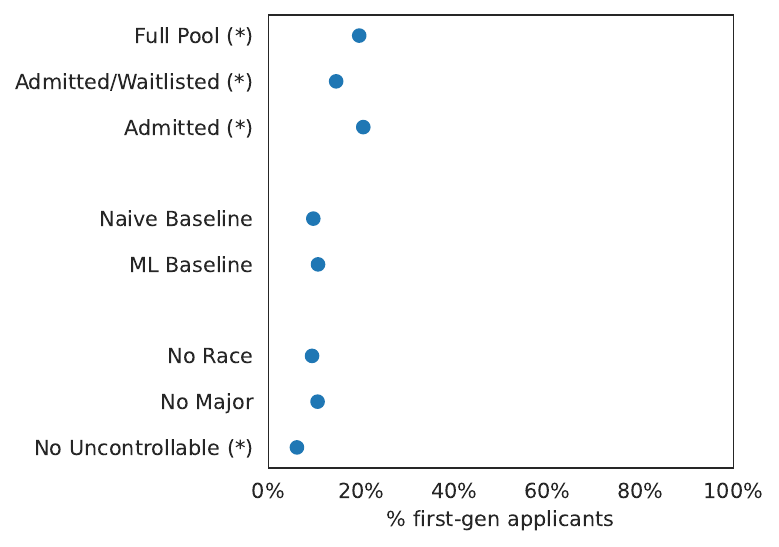}}
\hspace{0.05\textwidth}
\subfloat[]{\includegraphics[height=4.5cm]{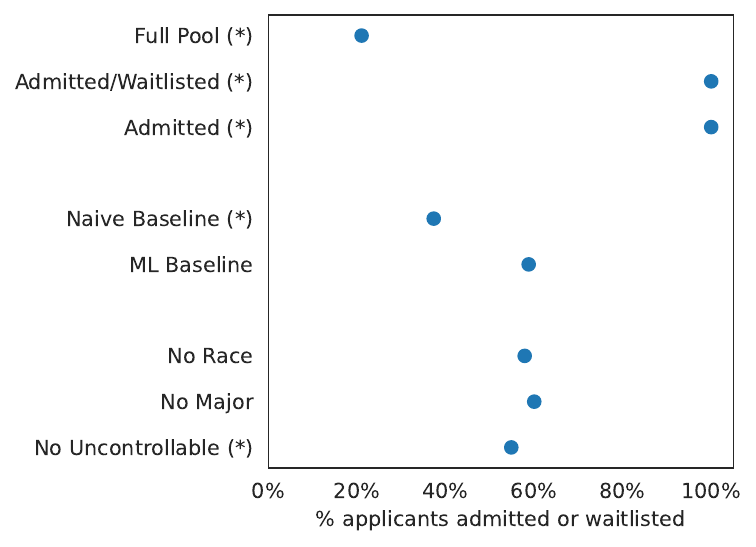}}
\hspace{0.05\textwidth}
\subfloat[]{\includegraphics[height=4.5cm]{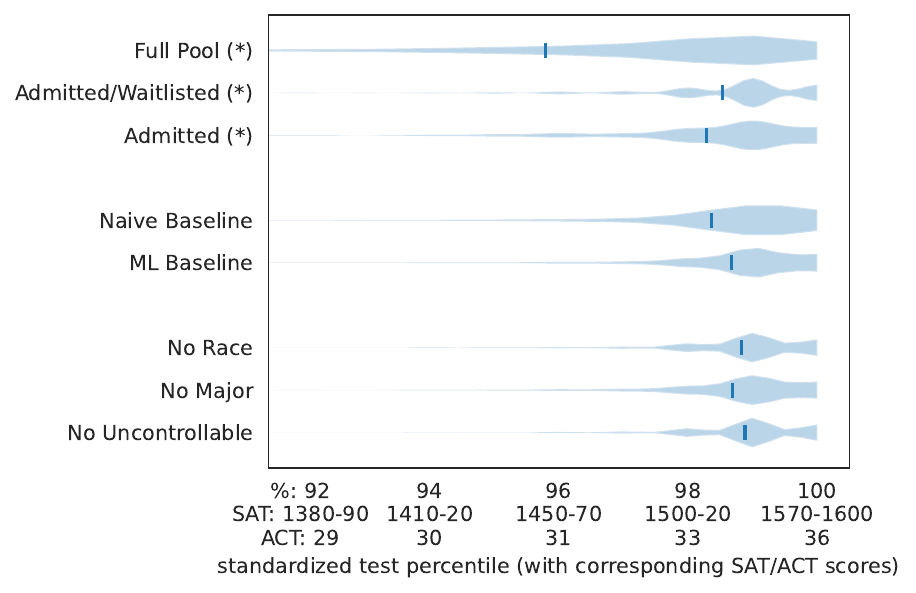}}
\Description{} 
\caption{Group outcomes when the ML ranking algorithms are trained to predict applicants' likelihood of being \textit{accepted or waitlisted} instead of only being \textit{accepted}.
\label{fig-appendix-target}}
\end{figure*}

The choice of target variable can have a major impact on the outputs of applicant ranking algorithms. To ensure that our results are not brittle with respect to our specific choice of target variable, we examined our group impact findings for both diversity and academic merit to determine whether our results still hold when ML ranking algorithms are trained to predict applicants' likelihood of being \textit{admitted or waitlisted} instead of simply being \textit{admitted}. This is an important choice, with potentially significant implications. The decision to admit an applicant in a holistic admissions process is based on a variety of qualitative and quantitative factors, including that applicant's background. To the extent that prior admissions decisions are based on affirmative action, it could arguably be a violation of the SFFA policy change to train admissions algorithms on prior admissions decisions, because such a practice would effectively encode and carry forward affirmative action. However, affirmative action may have less of an impact on who is placed on the waitlist. We see empirical evidence for this in Fig.~\ref{fig-appendix-target}(a), which shows that the racial demographics of applicants who are admitted or waitlisted are more similar to the demographics of the full applicant pool than the racial demographics of applicants who are admitted. Therefore, training ranking algorithms on waitlist decisions may represent a reasonable course of action for admissions offices, and we explore its impacts. 

As expected the ML baseline that is trained to identify admitted or waitlisted applicants includes a lower share of URM, low-income, and first-gen students in the top pool. It also includes in the top pool applicants with slightly higher standardized test scores. However, the trends discussed in \S\ref{s-results} still hold. Excluding race variables decreases the share of URM applicants in the top pool relative to the ML baseline Fig.~\ref{fig-appendix-target}(b); it also reduces the share of low-income applicants Fig.~\ref{fig-appendix-target}(c) but does not meaningfully change the share of first-gen applicants Fig.~\ref{fig-appendix-target}(d). At the same time, all ML models identify similar shares of applicants who were actually admitted or waitlisted in their top pools Fig.~\ref{fig-appendix-target}(e), and they also identify applicants with similarly high standardized test scores Fig.~\ref{fig-appendix-target}(f). 

\end{document}